\documentclass[12pt]{iopart}
\usepackage{bm}
\usepackage{amsopn}
\usepackage{amssymb}
\usepackage{amsfonts}
\usepackage{graphicx}
\usepackage[numbers,sort&compress]{natbib}
\usepackage{epstopdf}
\usepackage[pdftex]{color}
\usepackage{hyperref}
\usepackage{verbatim}
\usepackage{float}
\usepackage{textcomp}

\hypersetup{
    colorlinks=true,       
    linkcolor=cyan,          
    citecolor=magenta,        
    filecolor=magenta,      
    urlcolor=cyan,           
    runcolor=cyan
}

\newcommand{\beq}{\begin{eqnarray}}
\newcommand{\eeq}{\end{eqnarray}}
\newcommand{\bes} {\begin{subequations}}
\newcommand{\ees} {\end{subequations}}

\newcommand{\eqref}[1]{(\ref{#1})}
\newcommand{\cov}{\operatorname{cov}}

\newcommand{\ignore}[1]{}

\begin{document}

\title{Estimating the Density of States of Frustrated Spin Systems} 

\author{Lev Barash$^{1,2,3}$, Jeffrey Marshall$^4$, Martin Weigel$^5$, Itay Hen$^{4,6}$}

\address{$^1$ Landau Institute for Theoretical Physics, 142432 Chernogolovka, Russia}
\address{$^2$ Science Center in Chernogolovka, 142432 Chernogolovka, Russia}
\address{$^3$ National Research University Higher School of Economics, 101000 Moscow, Russia}
\address{$^4$ Department of Physics and Astronomy, and Center for Quantum Information Science \& Technology, University of Southern California, Los Angeles, California 90089, USA}
\address{$^5$ Applied Mathematics Research Centre, Coventry University, Coventry, CV1 5FB, United Kingdom}
\address{$^6$ Information Sciences Institute, University of Southern California, Marina del Rey, California 90292, USA}

\begin{abstract}

  Estimating the density of states of systems with rugged free energy landscapes is a
  notoriously difficult task of the utmost importance in many areas of physics
  ranging from spin glasses to biopolymers. Density of states estimation has also
  recently become an indispensable tool for the benchmarking of quantum annealers
  when these function as samplers. Some of the standard approaches suffer from a
  spurious convergence of the estimates to metastable minima, and these cases are
  particularly hard to detect.  Here, we introduce a sampling technique based on
  population annealing enhanced with a multi-histogram analysis and report on its
  performance for spin glasses. We demonstrate its ability to overcome the pitfalls
  of other entropic samplers, resulting in some cases in large scaling advantages
  that can lead to the uncovering of new physics. The new technique avoids some
  inherent difficulties in established approaches and can be applied to a wide range
  of systems without relevant tailoring requirements. Benchmarking of the studied
  techniques is facilitated by the introduction of several schemes that allow us to
  achieve exact counts of the degeneracies of the tested instances.

\end{abstract}
\maketitle

\section{Introduction} 
 
In statistical and condensed matter physics, the density of states (DOS) of a system
describes the number of states at each energy level. The DOS, which is independent of
temperature, represents a deep characterization of the system. In terms of
thermodynamics, knowledge of the DOS allows one to calculate the partition function
and hence all expectation values that can be derived from it, including the free and
internal energies as well as the specific heat, as a function of
temperature~\cite{sethna:06}. Alternatively, the DOS itself and closely related
quantities are the center of interest in an analysis of the thermodynamics of phase
transitions in the microcanonical ensemble~\cite{hueller:93}.  DOS calculations can
also determine the spacing between energy bands in
semiconductors~\cite{solidState1}.

Whilst  knowledge of the DOS, $\Omega(E)$, is extremely valuable, it cannot in
general be easily acquired. Exact calculations are only possible in a few special
cases such as Ising models on two-dimensional lattices
\cite{beale:96a,galluccio:00}. In general, the problem is exponentially hard as the
system size increases (in computational complexity terms it is a \#P-hard
problem)~\cite{DOScomplexity1,DOScomplexity2}. This difficulty notwithstanding, there
exist a number of approximation techniques, mostly based on Monte Carlo methods, that
allow one to estimate $\Omega(E)$. The most widely used approach of this type is the
Wang-Landau (WL) algorithm~\cite{wang:01a,wang:01b} and its variants
\cite{belardinelli:07,liang:07,vogel:13}, which is based on the multicanonical method
\cite{berg:92b}.

As stochastic approximation techniques, these approaches are affected by statistical
errors as well as systematic deviations (bias). Estimating statistical error is not
easily possible from a single WL simulation alone and normally requires statistics
over independent runs. The most relevant form of bias in WL simulations is that of a
false convergence, where the DOS estimate settles down on a deceptively smooth shape
that is however not a faithful representation of the actual DOS
\cite{schneider:17}. Naturally, such problems are notoriously hard to detect if no
independent information about the actual DOS is available.

Such difficulties apply in particular to systems with complex free-energy landscapes
that are typically accompanied by frustration in the interactions such as in the
protein-folding problem~\cite{bryngelson:95} and in the spin-glass systems that
result from a combination of frustration and quenched disorder~\cite{binder:86a}. The
latter may be viewed as prototypical classically-hard optimization problems, and they
are so challenging that specialized hardware has been built to simulate
them~\cite{belleti:09}. Recently, DOS estimation, which provides the relative
degeneracies of the energy levels of spin glasses, has become an indispensable tool
in the context of the benchmarking of experimental quantum
annealers~\cite{kadowaki_quantum_1998,farhi:01} and the attempts to demonstrate
speedups over classical devices.  The currently available commercial realization of
this paradigm~\cite{Dwave} effectively samples low-lying energy configurations of
spin-glass samples that are coded into the couplers connecting an array of
superconducting flux qubits. The properties of the resulting samples and the question
of whether such devices indeed provide superior performance as compared to classical
algorithms for some problem classes has been the subject of much recent
debate~\cite{hen:15,katzgraber:15,boixo:16,perdomo,adachi,Amin:boltzmann,fairInUnfair,exponentiallyBiased,uncertainFate,marshallrieffelhen:2017}.
The questions of reliable Boltzmann sampling aided by quantum annealers for machine
learning applications~\cite{perdomo,adachi,Amin:boltzmann} as well as the advantages
that quantum annealing devices potentially hold when tasked with fairly sampling the
ground-state manifolds of spin glasses with multiple minimizing
configurations~\cite{fairInUnfair,exponentiallyBiased,uncertainFate} are now a topic
of considerable interest.

Our goal in this work is threefold. i) We devise techniques for \emph{verifiably}
benchmarking algorithms for sampling the DOS, designed to overcome the
pitfalls of misinterpreting false convergences of entropic samplers. ii) Employing
the above techniques, we demonstrate the difficulties in applying traditional
algorithms for sampling the DOS to spin-glass instances. iii) We
introduce a population annealing algorithm for estimating the DOS that allows for the
intrinsic control of statistical and systematical errors and demonstrate how it can
outperform the standard approach without the associated problems of choosing energy
windows and related parameters that occur for the latter. 

\section{Verifiable benchmarking of entropic samplers\label{plantedSec}}

As mentioned above, approximation algorithms for the DOS are not always
reliable in converging to the correct answer, especially for the frustrated systems
considered here. While there are some general results regarding the convergence of
suitably modified WL type algorithms~\cite{liang:07}, convergence times can become
astronomically large and convergence hard to assess from intrinsic indicators
\cite{barash:17b}. As we shall see below, for disordered systems convergence times
can also fluctuate wildly between different realizations of the couplings. It is
hence highly desirable for benchmarking purposes to have at hand sets of samples that
are sufficiently challenging for the tested algorithms, but for which nevertheless
the (exact) DOS is known from other considerations. In general, such
samples are not readily available, but we present here two groups of examples that
are extremely useful in this respect: locally planar lattices and samples with
planted solutions.

For concreteness, we shall consider spin models of the Ising type, whose cost
function (or Hamiltonian) is of the form
\begin{equation}
  \label{eq:Hp}
  H = - \sum_{\langle i,j\rangle} J_{ij} s_i s_j - \sum_i h_i s_i,
\end{equation}
where $\langle i,j\rangle$ denotes the set of edges of the underlying graph.  Here,
$s_i=\pm 1$ are the Ising spin variables and the quenched parameters $J_{ij}$ and
$h_i$ denote the exchange couplings and random fields, respectively. In the
following, we will focus on the zero-field case $h_i = 0$. While the problems of
computing the DOS (or partition function) and of finding a ground state
for this problem in at least three dimensions are NP hard~\cite{barahona:82a}, both
tasks can be completed in polynomial time on any set of graphs of a genus bounded by
a constant, which includes planar and toroidal two-dimensional lattices
\cite{galluccio:00}. For such cases there exist efficient algorithms to solve the
above problems. For ground states, these include minimum-weight perfect matching
\cite{bieche:80a,gregor:07,khoshbakht:17a}, while for the partition function or
DOS the usual approaches are based on the counting of dimer coverings
which can be achieved via an evaluation of Pfaffians
\cite{blackman:91,saul:93,galluccio:00}.

Unfortunately, such techniques are restricted to locally planar graphs and so they do
not apply, for example, to the Chimera graphs used in current implementations of
quantum annealers, which have a genus that grows linearly in the number of sites. For
such more general problems we propose here an approach that is based on generating
problem instances for which the values and degeneracies of the ground- and first
excited-states, $\Omega(E_0)$ and $\Omega(E_1)$, are {\em exactly\/}
computable. Since the states of lowest energies are usually the most difficult to
sample, the degeneracies of these two energy levels are the most difficult to
ascertain, and a correct estimation of these serves as a good indicator of true
convergence. We create such samples by considering problem instances with {\em
  planted solutions}~\cite{hen:15,king:15,wang:17} --- an idea borrowed from
constraint satisfaction (SAT) problems~\cite{Barthel:2002tw,Krzakala:2009qo} where
the planted solution represents a ground-state configuration of Eq.~(\ref{eq:Hp})
that minimizes the energy and is known in advance. Following Ref.~\cite{hen:15}, the
Hamiltonian of a planted-solution spin glass is a sum of terms, each of which
consists of a small number of connected spins, namely, $H=\sum_j H_j$.  Each term
$H_j$ is chosen such that one of its ground states is the planted solution. It
follows then that the planted solution is also a ground state of the total
Hamiltonian. This class of instances has two attractive properties: i) the
ground-state energies of the generated problems are known in advance, and ii) the
\emph{exact} degeneracies of the ground and first excited states are
computable~\cite{fairInUnfair,marshallrieffelhen:2017}.  These in turn allow us to
check how close entropic samplers come to these exact values. The computation of
$\Omega(E_0)$ and $\Omega(E_1)$ is based on the fact that our generated instances
consist of a sum of terms, each of which has all minimizing configurations of $H$
as its ground state. To enumerate all ground states, we implement a form of the
`bucket' algorithm~\cite{dechter} designed to eliminate variables one at a time to
perform an exhaustive search efficiently (for a detailed description of the
algorithm, see the Supplementary Information of Ref.~\cite{fairInUnfair}). By noting
that the lowest energy excited states are those configurations that violate precisely
one clause, their degeneracy may also be calculated.  We perform the same
exhaustive search as above, but where now the configurations tested will correspond
to first excited states of one of the $H_j$ (and are still minimizing for
$H_{i \neq j}$). This gives the number of configurations which are ground states of
$H_{i \neq j}$ and first excited states of $H_j$; by
performing this calculation for each of the $H_j$, we get the total number of first
excited states of $H$.

While this approach in principle works for arbitrary graphs, we focus here on Chimera
lattices, i.e., two-dimensional arrays of unit cells of eight spins with a $K_{4,4}$
bipartite connectivity~\cite{Choi1,Choi2}, see for example
Ref.~\cite{marshallrieffelhen:2017}. Our choice is motivated by the attention these
graphs have gained in recent years in the context of optimization as well as sampling
via quantum annealing as the quantum annealers currently commercially available
feature qubits connected with this topology~\cite{johnson:11,berkley:13,Bunyk:2014hb}.
While the Chimera graph is
two-dimensional in nature~\cite{weigel:15}, it is also non-planar and as such gives
rise to difficult spin-glass problems~\cite{barahona:82}. We generated 625
planted-solution instances of 501 spins each\footnote{The 501 spins correspond to
  the graph considered in Ref.~\cite{marshallrieffelhen:2017}.}, following a technique
described in detail in Ref.~\cite{hen:15} wherein the clauses $H_j$ are chosen to be
`frustrated loops' along the Chimera graph. For each sample we employed the bucket
algorithm in order to obtain $\Omega(E_0)$ and $\Omega(E_1)$.

The combination of full exact DOS for samples on the square lattice and
toroidal boundary conditions and of exact values for $\Omega(E_0)$ and $\Omega(E_1)$
for the Chimera samples allows us to carefully examine the reliability and
performance of sampling schemes for estimating the DOS, avoiding the
pitfalls provided by badly converged estimates of stochastic approximation schemes.

\section{Sampling the DOS}

The common approximation algorithms for the DOS are based on Markov
chain Monte Carlo~\cite{berg:92b,wang:01a}.  In the following, we will use the most
popular of these, the Wang-Landau algorithm~\cite{wang:01a,wang:01b}, in a variant
dubbed the WL-$1/t$ method~\cite{belardinelli:07} that in principle can be shown to
converge to the correct answer if given infinite run time~\cite{liang:07}, as a
reference and contender of the method introduced here, entropic population annealing.

\subsection{Wang-Landau sampling}

In Wang-Landau (WL) sampling as introduced in Refs.~\cite{wang:01a,wang:01b} a
running estimate $\hat{\Omega}(E)$ of the DOS (initialized as
$\hat{\Omega}(E)=1$ $\forall E$) is updated in a random walk through energy space by
multiplying $\hat{\Omega}(E)$ at the current energy $E$ by a modification factor $f$
(initially chosen to equal Euler's number $e$) at each step. A new configuration of
energy $E'$ is proposed according to the chosen move scheme (for a spin model
typically through single spin flips) and accepted with probability
\begin{equation}
  p_\mathrm{acc} = \min\left[1,\frac{\hat{\Omega}(E)}{\hat{\Omega}(E')}\right].
  \label{eq:wlprob}
\end{equation}
If, after some time, the histogram $\hat{H}(E)$ of all possible energies is found to
be ``sufficiently flat'' (typically interpreted as no histogram bin having less than
80\% of the average number of entries~\cite{wang:01a}, but see also
Ref.~\cite{gross:17} for a related discussion), the modification factor is reduced as
$f \to \sqrt{f}$, and the histogram $\hat{H}(E)$ is reset to an empty state. The
algorithm stops if $f$ is ``sufficiently small'', for example
$f=f_\mathrm{final}=\exp(10^{-8})$. While the approach was invented as a variant of
Markov chain Monte Carlo, the fact that the transition probabilities according to
Eq.~(\ref{eq:wlprob}) change constantly means that neither detailed nor global
balance are satisfied, and it is more useful to think of the method as a ``stochastic
approximation algorithm''~\cite{liang:07}.

It is well known that the original scheme of Ref.~\cite{wang:01a,wang:01b} does not
converge to the true DOS, but the error asymptotically saturates at a value
determined by the protocol used for reducing $f$~\cite{qiliang:03}. This shortcoming
is remedied by choosing a different modification protocol for $f$, leading to a
slower decay of $f$ at late times. The so-called $1/t$ algorithm proposed in
Ref.~\cite{belardinelli:07} uses two phases. In the first phase the standard WL
algorithm is used, with the only difference that the energy histograms are considered
to be sufficiently flat already if $\hat{H}(E)\ne 0$ for all $E$. Once $\ln f$ falls
below the moving threshold $N_E/t$, where $t$ is the simulation time measured in
spin-flip attempts and $N_E$ is the number of energy levels, the simulation enters
the second phase. There, the modification factor is adapted quasi continuously
according to $\ln f(t) = N_E/t$ and histogram flatness is no longer tested. The
simulation stops once $f(t) \le f_\mathrm{final}$.

While no saturation of error occurs in the $1/t$ algorithm
\cite{belardinelli:07,schneider:17}, it is still necessary to know the permissible
energy levels (including the ground state) beforehand to judge histogram flatness,
which is a major drawback of the method for disordered systems. In practice, we
therefore employ pre-runs of the WL type without any reduction of the modification
factor with the goal of discovering the available energy levels\footnote{We generally
  choose the length of pre-runs so as to make sure that all levels are discovered,
  but in a few cases the actual ground state is only found in the main run.}. For
large systems and problems with complex free-energy landscapes, it is usually
necessary to divide the total energy range into several windows for which the
algorithm is run separately to achieve convergence on realistic time scales for
interesting system sizes \cite{wang:01b}. The right choice of window sizes in such
schemes is a difficult problem especially for disordered and frustrated
systems~\cite{yin:12}, and we are not aware of any reliable systematic approach to
solve it. As a consequence, we had to spend considerable time with trial and error to
arrive at suitable setups for the problems studied below. A number of further
generalizations of the method have been proposed, for instance a combination with
parallel tempering~\cite{vogel:13} which uses progressively smaller windows at lower
energies, but here again there is no general algorithm for determining the
appropriate window sizes automatically.

\subsection{Entropic population annealing}

The new algorithm introduced here, which we call entropic population annealing (EPA),
is not based on Markov chains but on the sequential Monte Carlo method. Population
annealing (PA) was first studied in Refs.~\cite{iba:01,hukushima:03} and more recently
developed further in
Refs.~\cite{machta:10a,wang:15a,borovsky:16,barash:16,barash:17,amey:18,barzegar:17}. It
is based on the initialization of a population of replicas drawn from the equilibrium
distribution at high temperatures, which is then subsequently cooled to lower and
lower temperatures. During this process, a combination of population control and spin
flips is used to ensure that the ensemble remains in equilibrium. The simulation
entails the following steps~\cite{machta:10a,barash:16}:
\begin{enumerate}
\item Set up an equilibrium ensemble of $R_0 = R$ independent copies (replicas) of
  the system at inverse temperature $\beta_0=1/k_BT_0$.
\item Take a step to inverse temperature $\beta_i > \beta_{i-1}$ by resampling the
  configurations $j = 1,\ldots, R_{i-1}$ with their relative Boltzmann weight
  $\hat{\tau}_i(E_j)$, leading to $R_i \ne R_{i-1}$ replicas, in general.
\item Update each replica by $\theta$ rounds of an MCMC algorithm at inverse
  temperature $\beta_i$.
\item Goto step 2 unless the inverse target temperature $\beta_\mathrm{f}$ has been
  reached.
\end{enumerate}
During resampling, the expected number of copies is
\begin{equation}
  \hat{\tau}_i(E_j) = \frac{R}{R_{i-1}}\frac{e^{-(\beta_i-\beta_{i-1})E_j}}{Q(\beta_{i-1},\beta_i)},
\end{equation}
with a  normalizing factor
\begin{equation}
  Q(\beta_{i-1},\beta_i) = \frac{1}{R_{i-1}}
  \sum_{j=1}^{R_{i-1}} e^{-(\beta_i-\beta_{i-1})E_j}.
  \label{eq:Q}
\end{equation}
The actual number of copies is taken to be the integer part
$\lfloor\hat{\tau}_i(E_j)\rfloor$ plus an additional copy added with a probability
corresponding to the fractional part,
$\hat{\tau}_i(E_j) - \lfloor\hat{\tau}_i(E_j)\rfloor$. While initially constant
(inverse) temperature steps were used on increasing $\beta_i > \beta_{i-1}$
\cite{machta:10a}, it turns out that a better, parameter-free method consists of
choosing $\beta_i$ to ensure a certain overlap of the energy distributions between
the two temperatures~\cite{barash:16}. This overlap can be computed from the
resampling factors,
\[
  \alpha(\beta_{i-1},\beta_i) = \frac{1}{R_{i-1}}\sum_{j=1}^{R_{i-1}} \min\left(1,
    \frac{R\exp[-(\beta_i-\beta_{i-1})E_j]}{R_{i-1}Q(\beta_{i-1},\beta_i)}
  \right),
\]
and $\beta_i$ is adapted using a bisection search such as to ensure an overlap
$\alpha^\ast$ of energy histograms. The method is not very sensitive to the precise
value of $\alpha^\ast$, and we choose $\alpha^\ast = 0.86$ in the runs below.

While the algorithm described above is just population annealing~\cite{machta:10a}
improved by adaptive temperature steps~\cite{barash:16,amey:18}, the possibility of
sampling the entropy arises from a combination of the method with multi-histogram
techniques~\cite{ferrenberg:89a}. An estimator of the free energy follows directly
from the resampling factors~\cite{machta:10a},
\begin{equation}
  -\beta_i \hat{F}({\beta_i}) = \ln Z_{\beta_0} + \sum_{k=1}^{i} \ln Q_k,
  \label{eq:free-energy}
\end{equation}
where $Z_{\beta_0}$ is the partition function at the initial temperature
$\beta_0$. In the following, we always choose $\beta_0 = 0$, such that simply
$Z_{\beta_0}=2^N$, where $N$ is the number of spins.  We can then estimate the
DOS by combining the histograms at all temperature steps. As we show in
\ref{sec:dos-estimator}, a variance-optimized estimator is given by
\begin{equation}
  \hat{\Omega}(E) = \frac{\sum\limits_{i = 1}^{N_\beta} \hat{H}_{\beta_i}(E) }{\sum\limits_{i = 1}^{N_\beta} R_i
    \exp[\beta_i \hat{F}(\beta_i)-\beta_iE]}.
  \label{eq:MHReq1}
\end{equation}
Here, $N_\beta$ is the total number of temperatures, and the energy histogram
$\hat{H}_{\beta_i}(E)$ at inverse temperature $\beta_i$ is normalized such that
$\sum_E \hat{H}_{\beta_i}(E) = R_i$. In Eq.~\eqref{eq:MHReq1}, the free-energy
estimate $\hat{F}(\beta_i)$ is taken from Eq.~\eqref{eq:free-energy}. More
sophisticated estimators that can lead to improved results in some cases are
discussed in \ref{sec:dos-estimator}.

This approach is naturally suited for (moderately or massively) parallel calculations
as the $R$ replicas are simulated independently of each other and the only
interaction occurs during resampling. An efficient GPU implementation was discussed
in Ref.~\cite{barash:16}. Importantly for our application, EPA does not require any
prior knowledge of the range of realized energies. Additionally, as we shall see
below, EPA performs better at estimating $\Omega(E)$ for hard spin-glass samples than
the WL-$1/t$ algorithm.  A detailed analysis of systematic and statistical errors of
PA can be found in Ref.~\cite{weigel:17a}. Here it is worthwhile to note that
statistical errors can be estimated from a single run by a jackknife blocking
analysis over the population that is introduced in Ref.~\cite{weigel:17a}. This is
further discussed in \ref{sec:error-appendix}. Also, note that it is possible to
include histograms from independent runs in the overall estimate provided through
Eq.~\eqref{eq:MHReq1} by extending the sums over $i$ over the temperature steps of
all runs. The relative weight of these contributions is automatically taken into
account through the free-energy factors deduced from Eq.~\eqref{eq:free-energy} and
the population sizes $R_i$. This is a natural generalization of the weighted averages
first proposed for more basic observables in Ref.~\cite{machta:10a}. This scheme
makes it possible to determine the DOS with arbitrary accuracy in a fixed time given
sufficient parallel resources.

It should be noted that, as it stands, EPA only visits energies in the physical
region $E\lesssim 0$, which is in contrast to WL that naturally also explores
energies $E > 0$. Should one be interested in this unphysical regime, however, it is
possible to derive its DOS from EPA, too. For systems on bipartite graphs the DOS is
always symmetric, $\Omega(-E) = \Omega(E)$, so it is easy to construct the full DOS
while just actually sampling energies $E \le 0$. This is the case for all examples
discussed below. For more general situations, it is also possible to construct the
full DOS by running EPA twice, once for the Hamiltonian $H$ and once for $-H$ and
combining the results.

\section{Results}

In order to test the efficiency of EPA against the WL-$1/t$ algorithm in an objective
way that is unaffected by problems of false convergence, we applied the two
algorithms to the planted solutions on Chimera graphs as well as to the stochastic
$\pm J$ model on the square lattice with periodic boundaries, for both of which we
have exact results. As a baseline for the comparison, we tested both methods for the
case of the Ising ferromagnet on a square lattice for which extensive exact results
are available. There, we find similar performance of the two techniques, see the
discussion in~\ref{sec:ising}.

\subsection{Ising spin glasses on Chimera graphs}
\label{sec:planted}

We first considered the Chimera spin-glass instances with planted solutions and
$N=501$ spins. In order to be able to compare the two algorithms on an equal footing,
we could have directly considered them to be allowed the same runtime. This measure,
however, is implementation and platform specific.  Since both algorithms spend most
of their time flipping spins, we compare them for simulations employing the same
number $2\times 10^{12}$ of spin-flip attempts. For the used (serial) code for WL-$1/t$ this
corresponded to a wallclock time of 37~h (on an Intel Xeon 2.4 GHz CPU), while for
EPA we used a massively parallel GPU program~\cite{barash:16} that took approximately
1.2~h (on an Nvidia Tesla K40 GPU) per realization.

As mentioned above, for WL it is required to know the allowed energy levels to decide
about the flatness of histograms. This knowledge was here acquired by a pre-run of
the WL type employing $2\times 10^{11}$ spin-flip attempts and with a fixed
modification factor $\ln f = 1$ to explore the energy landscape (the corresponding
run-time is included in the time estimate given above). This knowledge is not
required for the EPA runs. With a single window covering the full energy range,
WL-$1/t$ did not complete phase 1 for the vast majority of samples. We therefore used
two windows with energy ranges $[E_0,E_0+1200]$ and $[E_0+1100,50]$ in dimensionless
units, respectively. The spin-flips were divided evenly between the two energy windows.
Here, the disorder average of the ground state energy $E_0$ was found to be $-3635$.
The energy levels were determined during the
pre-run, which found the ground state in the vast majority of cases. It is clear that
for larger systems where it is much harder to find the ground state the determination
of suitable windows for WL-$1/t$ becomes much harder. The simulation was started in a
random configuration within the energy range of the window. With that restriction,
$567$ out of $625$ samples completed phase 1 in the first window 
within the remaining $8\times 10^{11}$ flip attempts after the pre-run. 
No range restriction was required for EPA, and we used a population of size
$R=3\,992\,000$ with $\theta=10$ rounds of spin flips per resampling step and a
histogram overlap $\alpha^\ast = 0.86$, resulting in typically 100 temperature steps
down to $\beta_\mathrm{f} = 5\times 10^4$\footnote{The unusual population size
  results from the attempt of matching the average number of spin-flip attempts
  between the two methods.}.
 
\begin{figure}[tb!] 
\begin{center}
  \includegraphics[width=0.7\textwidth]{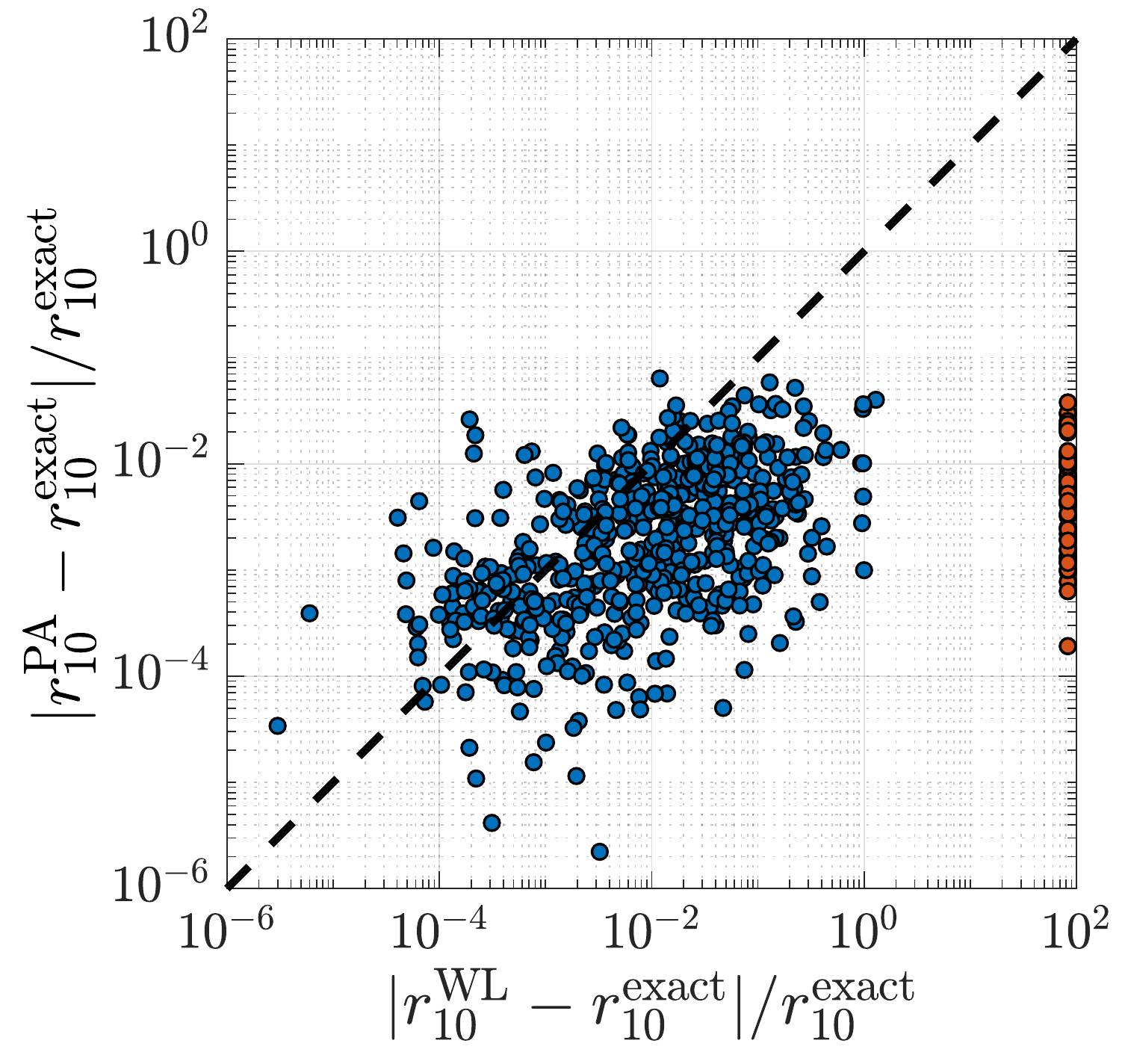}
\end{center}
\caption{ Scatter plot of the relative error of the WL-$1/t$ and the EPA algorithms
  in estimating the ratio $r_{10} = \Omega(E_1)/\Omega(E_0)$ of degeneracies for
  $N=501$ Chimera spin-glass samples with planted solutions.  Both algorithms applied
  a total of $2\times 10^{12}$ spin-flip attempts per sample, including an additional
  pre-run for WL-$1/t$ to determine the allowed energy values. 
  For 54 out of 625 samples the deviation for WL-$1/t$ falls outside the
  range of the plot, and these are shown at the right edge of the plot in red.  }
  \label{fig:chimeraPAvsWL}
\end{figure}
	
We note that both EPA and WL-$1/t$ intrinsically estimate entropy {\em
  differences\/}, i.e., ratios of degeneracies for neighboring energy values, and the
absolute scale is only achieved through an additional normalization such as that
given by $Z_{\beta_0}$ in Eq.~(\ref{eq:free-energy}). It is therefore reasonable to
study their performance in estimating
\[
  r_{10} = \frac{\Omega(E_1)}{\Omega(E_0)},
\]
the ratio of degeneracies of first excited and ground states.  In
Fig.~\ref{fig:chimeraPAvsWL} we show the relative deviations of the ratios $r_{10}$
from the exact values known through the planting, as estimated from WL-$1/t$ and
EPA. WL-$1/t$ found the correct ground-state energy for $622$ of the $625$
samples. For some samples the relative deviations are so large that they exceed the
scale of the plot of Fig.~\ref{fig:chimeraPAvsWL}, some by many orders of
magnitude\footnote{This includes the three samples for which WL-$1/t$ does not find the
  ground state, which effectively implies that $r_{10}$ and the relative deviation
  are infinite.}. These samples are shown at the boundary of the box and in a
different color. It is clear that for most samples the deviations are substantially
smaller for EPA than for WL-$1/t$. In total, EPA outperforms WL-$1/t$ in 80\% of the
instances.  The error of WL-$1/t$ is larger than 7\% for 25\% of samples and it is
difficult to distinguish between the accurate and inaccurate WL results.  In
contrast, the EPA results are accurate to within 7\% for all of the $625$ samples.

\subsection{Ising spin glasses on toroidal graphs}

\begin{figure}[tb!] 
\begin{center}
  \includegraphics[width=0.7\textwidth]{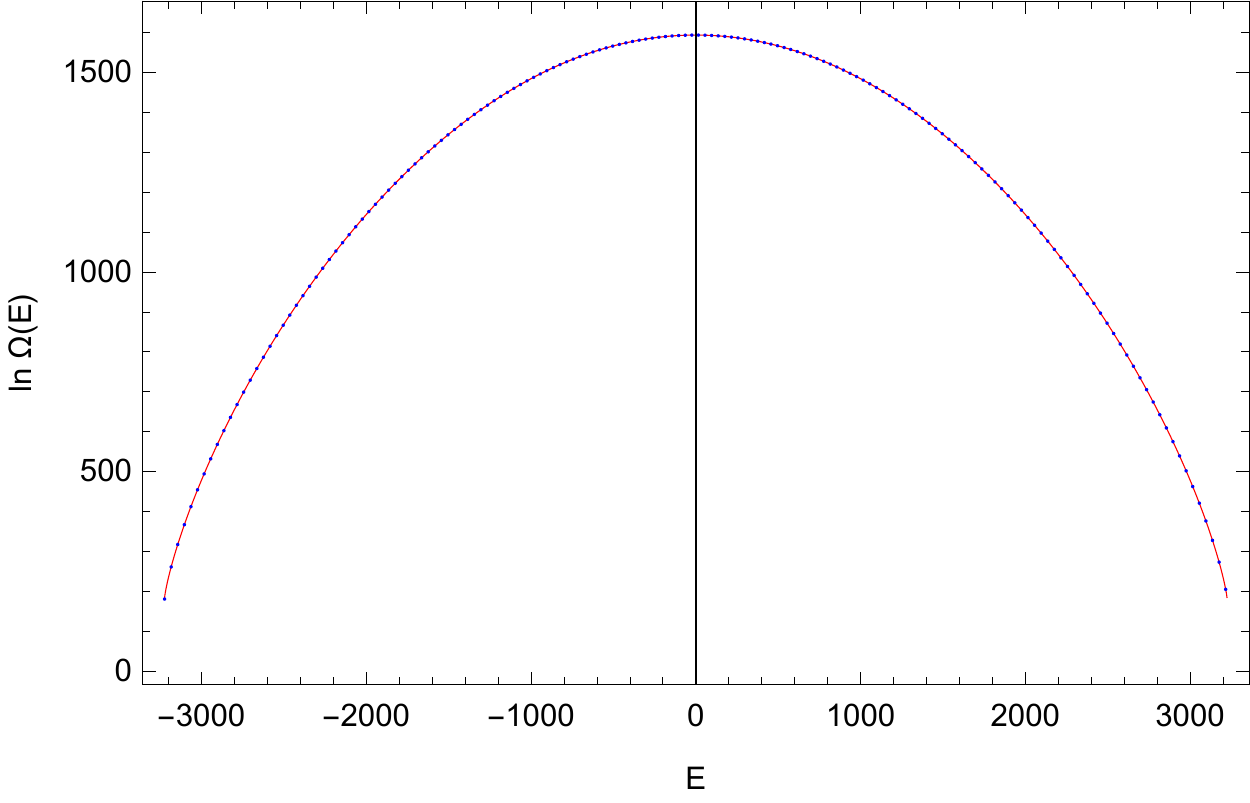}
  \includegraphics[width=0.7\textwidth]{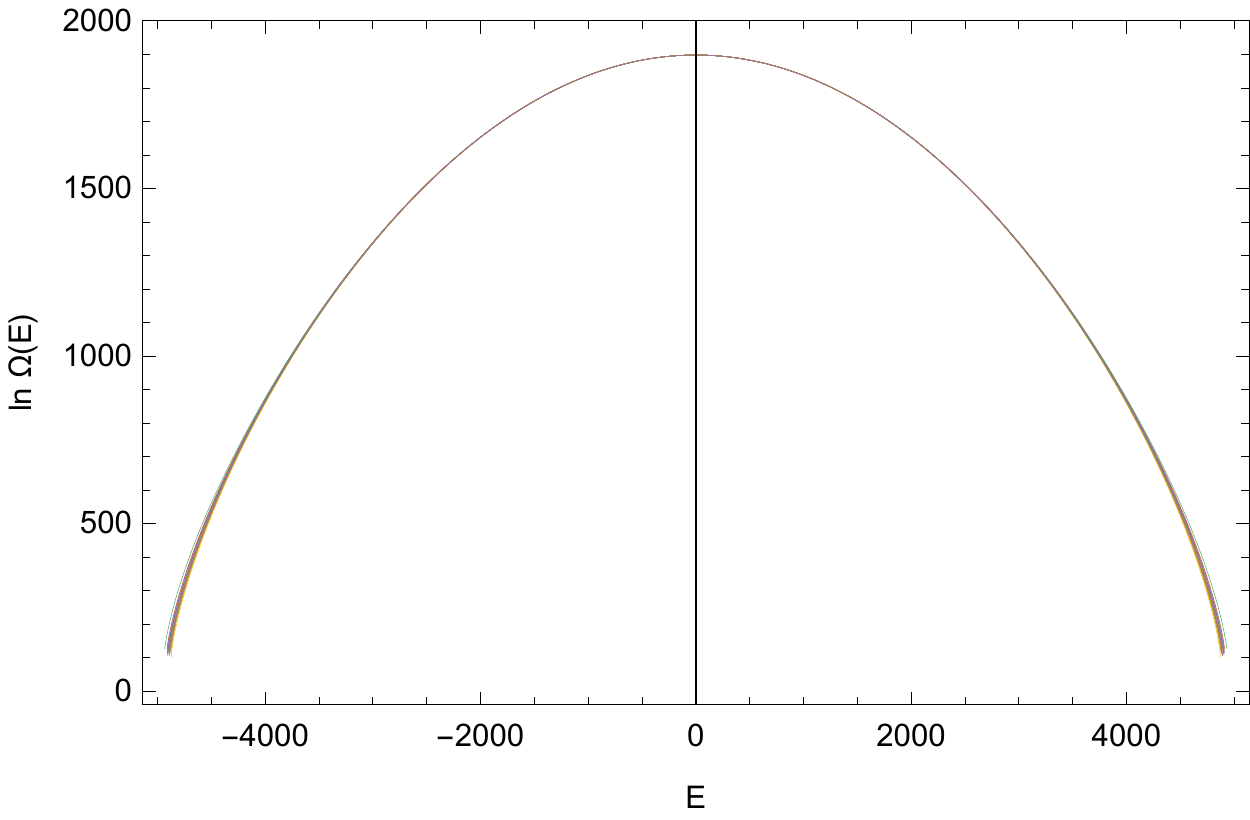}
\end{center}
\caption{ Examples of the DOS for the Edwards-Anderson spin glass. Top: Results from
  EPA runs for a single $L=48$ toroidal lattice sample (points) as compared to the
  exact DOS calculated from the Pfaffian method. Bottom: DOS estimated from EPA runs
  for 20 $L=14$ 3D $\pm J$ samples.  }
  \label{fig:DOS-examples}
\end{figure}

For planar or otherwise two-dimensional lattices of a fixed genus, a counting of
dimer coverings and the corresponding evaluation of Pfaffians can be used to
determine the full DOS in polynomial time
\cite{blackman:91,saul:93,galluccio:00}. We studied toroidal graphs, i.e., $L\times L$
patches of the square lattice with periodic boundary conditions using the
implementation proposed in Ref.~\cite{galluccio:00} which has an asymptotic run-time
scaling of O($L^5$). Using this approach, we evaluated $1000$ samples with a standard
$\pm J$ coupling distribution and $32\times 32$ spins, and also $500$ samples of size
$48\times 48$. An example of $\ln \Omega(E)$ as estimated for a single sample of size
$L=48$ from EPA is shown in the top panel of Fig.~\ref{fig:DOS-examples}. At this
scale, the data are completely indistinguishable from the exact result also shown for
comparison. As one reads off from the graph, the actual DOS $\Omega(E)$ spans about
700 orders of magnitude, and it is quite remarkable that it can be estimated so
accurately from the simulations. To systematically assess the accuracy of the
sampling for different disorder realizations, we considered the total deviation $\Delta$
of the simulation results from the exact DOS, where
\begin{equation}
  \Delta = \frac{1}{N_E}\sum_{i=1}^{N_E} \left|\frac{\ln \Omega(E_i)-
      \ln \Omega_{\mathrm{exact}}(E_i)}{\ln \Omega_{\mathrm{exact}}(E_i)} \right|.
  \label{eq:delta}
\end{equation}
While for PA an absolute normalization of $\Omega(E)$~\cite{weigel:02a} follows from
the free-energy estimator Eq.~(\ref{eq:free-energy}) in combination with
Eq.~(\ref{eq:MHReq1}), WL-$1/t$ as described above only yields the DOS up to an
overall factor. To fix the latter we used the fact that the sum $\sum \Omega(E)$ over
all energy levels must equal the total number $2^N$ of states. Note that different
ways of normalization of WL-$1/t$ lead to quite different fluctuations of the final
DOS estimates, and the normalization via the total number of states used here leads
to the best results, see the discussion in~\ref{sec:normalization}. Since EPA
only samples states with energies $E \lesssim 0$, we restricted the energy range for
WL-$1/t$ to $E\le 50$ to ensure a fair comparison\footnote{For the normalization of
  the DOS we hence completed $\Omega(E) = \Omega(-E)$ for the positive energies. For
  consistency, we additionally applied the same normalization in EPA.}. 

For $32\times 32$ samples, we used $1.8\times 10^{12}$ spin-flip attempts in the main
run of WL-$1/t$ employing a single energy window with $E\le 50$.  Just as for the Chimera
samples, a pre-run was required to determine the range of possible energies, for
which an additional $2\times 10^{11}$ updates were used. For the EPA algorithm, we
used a population size $R=2\,340\,000$, and performed $\theta=19$ rounds of
spin-flips between two resampling steps. The imposed histogram overlap of
$\alpha^\ast = 0.55$ resulted in $N_\beta= 44$ temperature steps for most disorder
realizations down to $\beta_f = 5\times 10^4$. The total number of spin flips in
these EPA runs is hence also approximately $2\times 10^{12}$.  For system size
$48\times 48$, WL-$1/t$ required two energy windows to converge; these were chosen as
$[E_0,E_0+64]$ and $[E_0+36,50]$. After the pre-run of length $6\times 10^{11}$
across both windows, we used $3.1\times 10^{12}$ spin-flip attempts in the main run
of the first window and $1.0\times 10^{12}$ updates in the second window. The two DOS
segments obtained by WL-$1/t$ were sewn together by matching the estimates at a point
in the intersection of the two windows. For the EPA algorithm, we used
$R=1\,019\,965$, $\theta=10$ and $\alpha^\ast = 0.86$, which resulted in
$N_\beta = 200$ ($\beta_\mathrm{f} = 3$) and hence the total number of spin flips is
$4.7\times 10^{12}$ as for the WL-$1/t$ runs.

\begin{figure}[tb!] 
\begin{center}
  \includegraphics[width=0.55\textwidth]{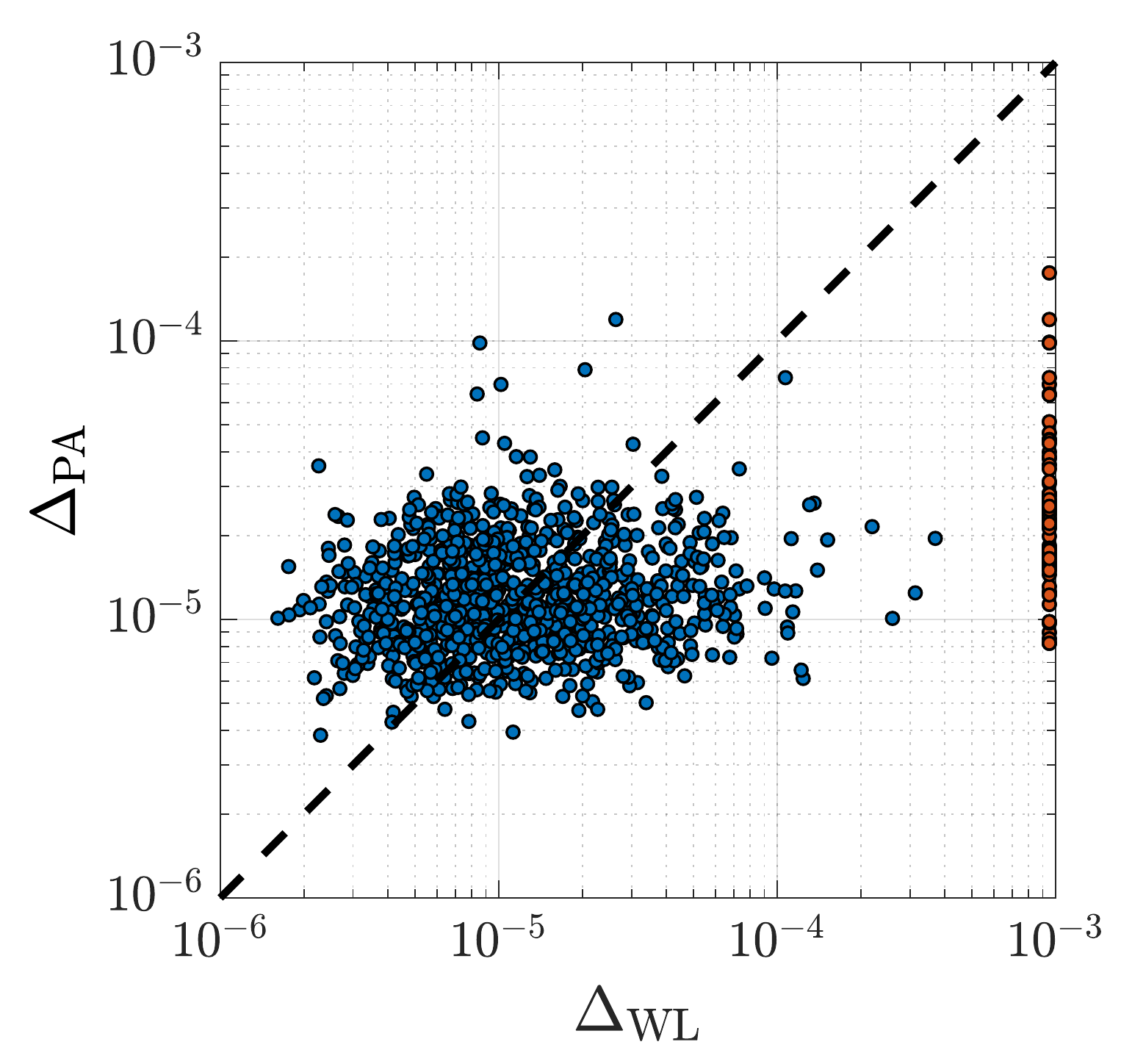}
  \includegraphics[width=0.55\textwidth]{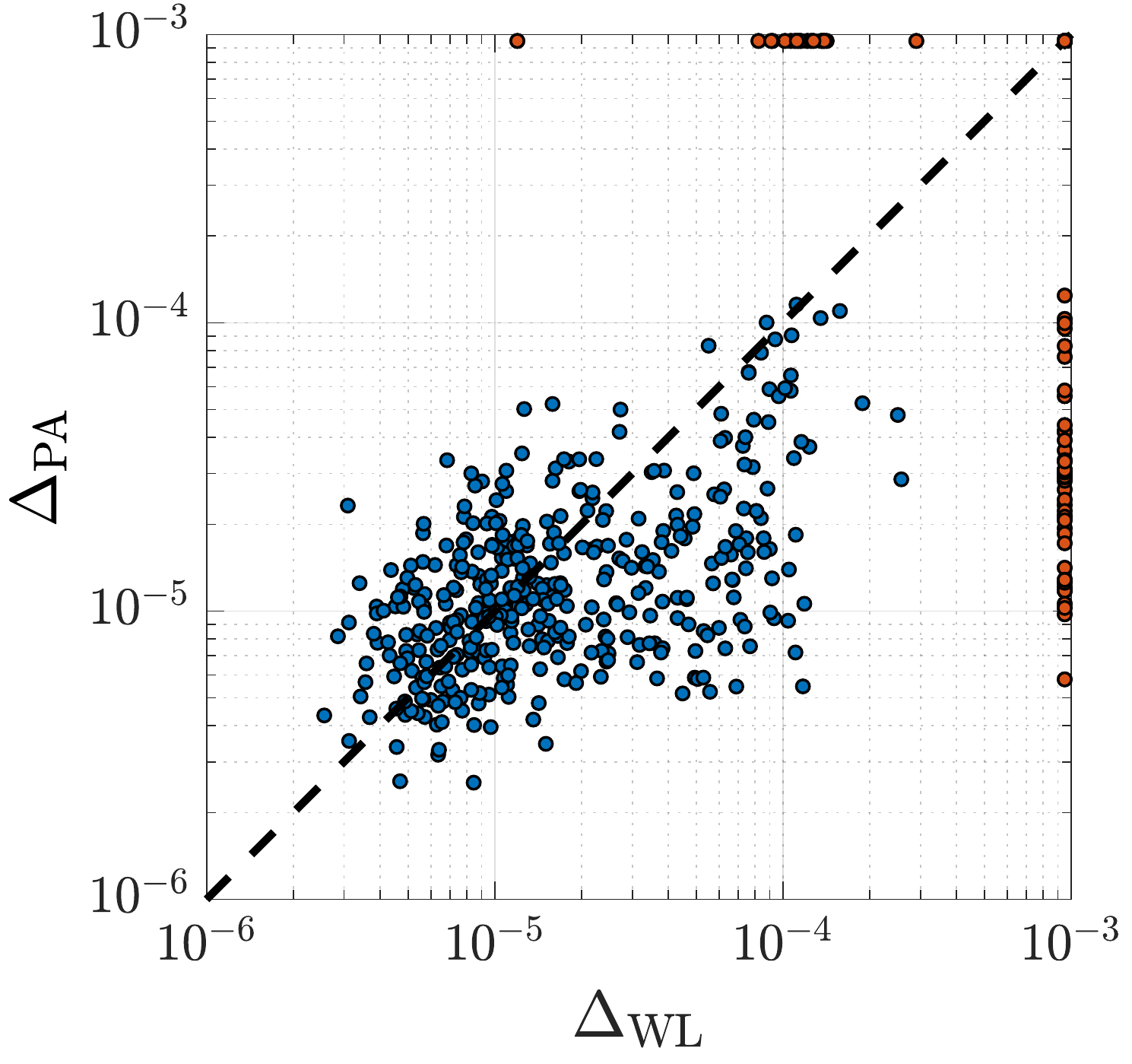}
\end{center}
\caption{Scatter plots of the average relative deviation of level entropies from the
  exact result according to Eq.~(\ref{eq:delta}) for toroidal $\pm J$ spin-glass
  samples of size $L\times L$ spins as estimated by the WL-$1/t$ and EPA methods. Top
  panel: $L=32$. Both algorithms were run for a total of $2\times 10^{12}$ attempted
  spin flips per sample. EPA outperforms WL-$1/t$ for 52\% of instances (516 out of
  1000). For 79 samples the deviations for WL-$1/t$ fall outside the scale of the
  plot and these samples are drawn at the right edge, in red.  Bottom panel:
  $L=48$. Both algorithms were run for a total of $4.7\times 10^{12}$ attempted spin
  flips per sample. EPA outperforms WL-$1/t$ for 62\% of instances (312 out of
  500).}
  \label{fig:toroidalPAvsWL}
\end{figure}

The resulting values for the average relative deviation of the level entropies
according to Eq.~(\ref{eq:delta}) are shown in the scatter plots provided in
Fig.~\ref{fig:toroidalPAvsWL}.  The top panel corresponds to 1000 samples of size
$32\times 32$. In some cases the ground states were not found or the first phase of
WL-$1/t$ did not complete, leading to extremely large or infinite deviations; the
corresponding samples are shown in red at the boundary of the plot. In this case we
only find a moderate advantage for EPA, which outperforms WL-$1/t$ on  516 of the
1000 samples. Considering the larger system size $L=48$, the advantage of EPA
increases, leading to a smaller value of $\Delta$ for  291 samples out  of the 451
samples where both methods found all energy levels. This observation is in line with
a general trend of EPA faring relatively better for harder problems as compared to WL
that we shall see confirmed for other examples below.

\subsection{Three-dimensional Ising spin glasses}
\label{sec:3d_samples}

Concerning the trend of improving relative performance of EPA for harder problems, it
is interesting to see how the two samplers perform on spin-glass instances in three
dimensions, where the spin-glass problem is known to be NP-hard
\cite{barahona:82a}. To this end, we studied samples of the $\pm J$ Edwards-Anderson
model with $L=8$ and $L=14$. For $L=8$ we were able to employ a single energy window
for WL, and the parameters for EPA were $R=1\,992\,984$, $\theta = 10$,
$\alpha^\ast = 0.86$, and $\beta_\mathrm{f} = 3$, using $10^{12}$ spin-flip attempts
in both cases. For $L=14$, $5.5\times 10^{12}$ spin flips were applied in each
run. For the WL-$1/t$ method we used two windows, $[E_0, E_0 + 64]$ and
$[E_0 + 36, 50]$, with $3.9\times 10^{12}$ and $1.0\times 10^{12}$ in the main run,
respectively. The remaining $6\times 10^{11}$ spin-flip attempts were used in the
pre-run. The parameters for EPA were $R = 867\,694$, $\theta = 10$,
$\alpha^\ast = 0.86$, and $\beta_\mathrm{f} = 10$. In the bottom panel of
Fig.~\ref{fig:DOS-examples} we show $\ln \Omega(E)$ for the $L=14$ samples. The
sample-to-sample fluctuations in the DOS are in fact rather small and can only be
seen in the very low-energy part of the spectrum (as well as its mirror image for
large positive energies).

\begin{figure}[!tb] 
\begin{center}
\includegraphics[width=0.7\textwidth]{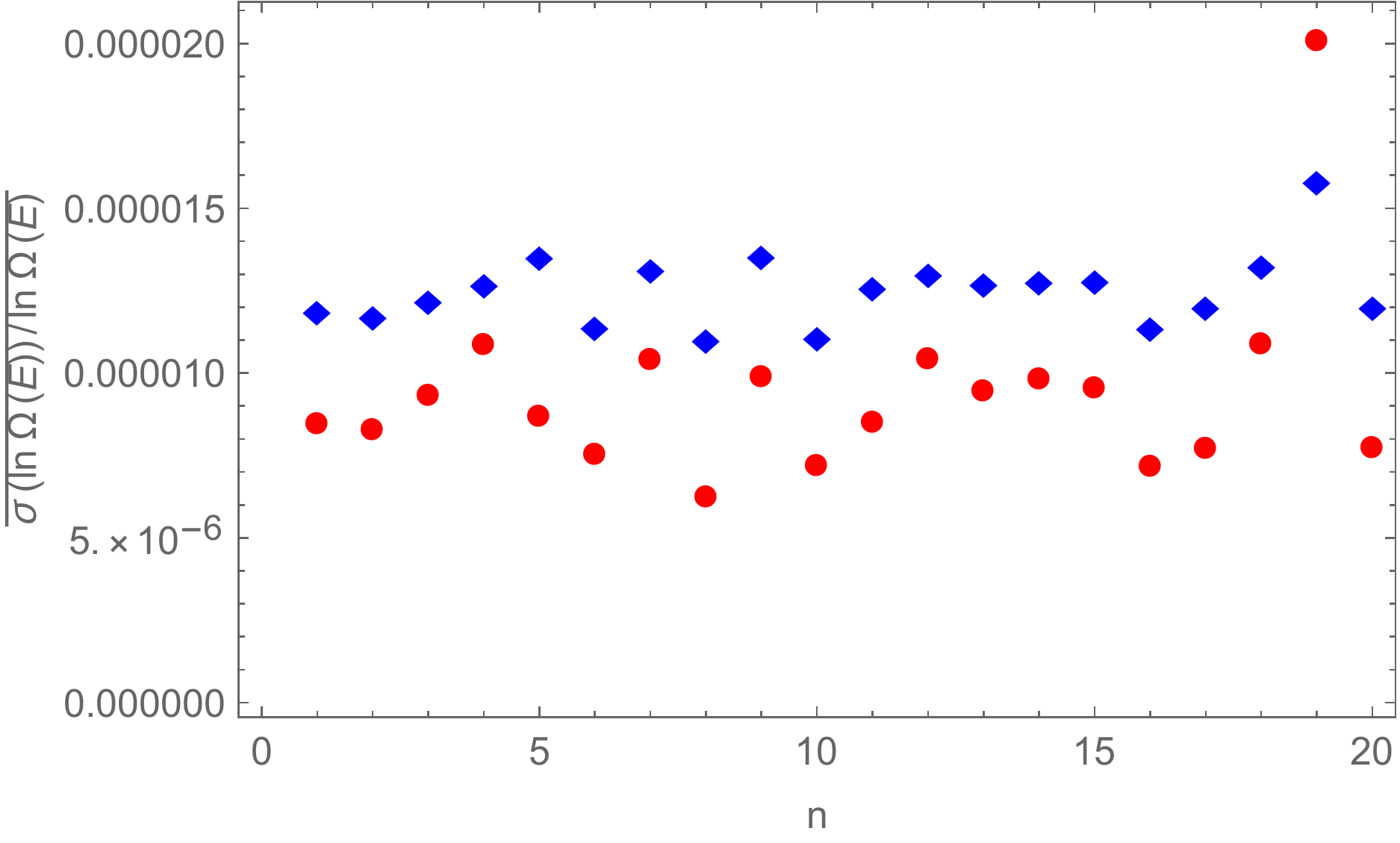}
\includegraphics[width=0.7\textwidth]{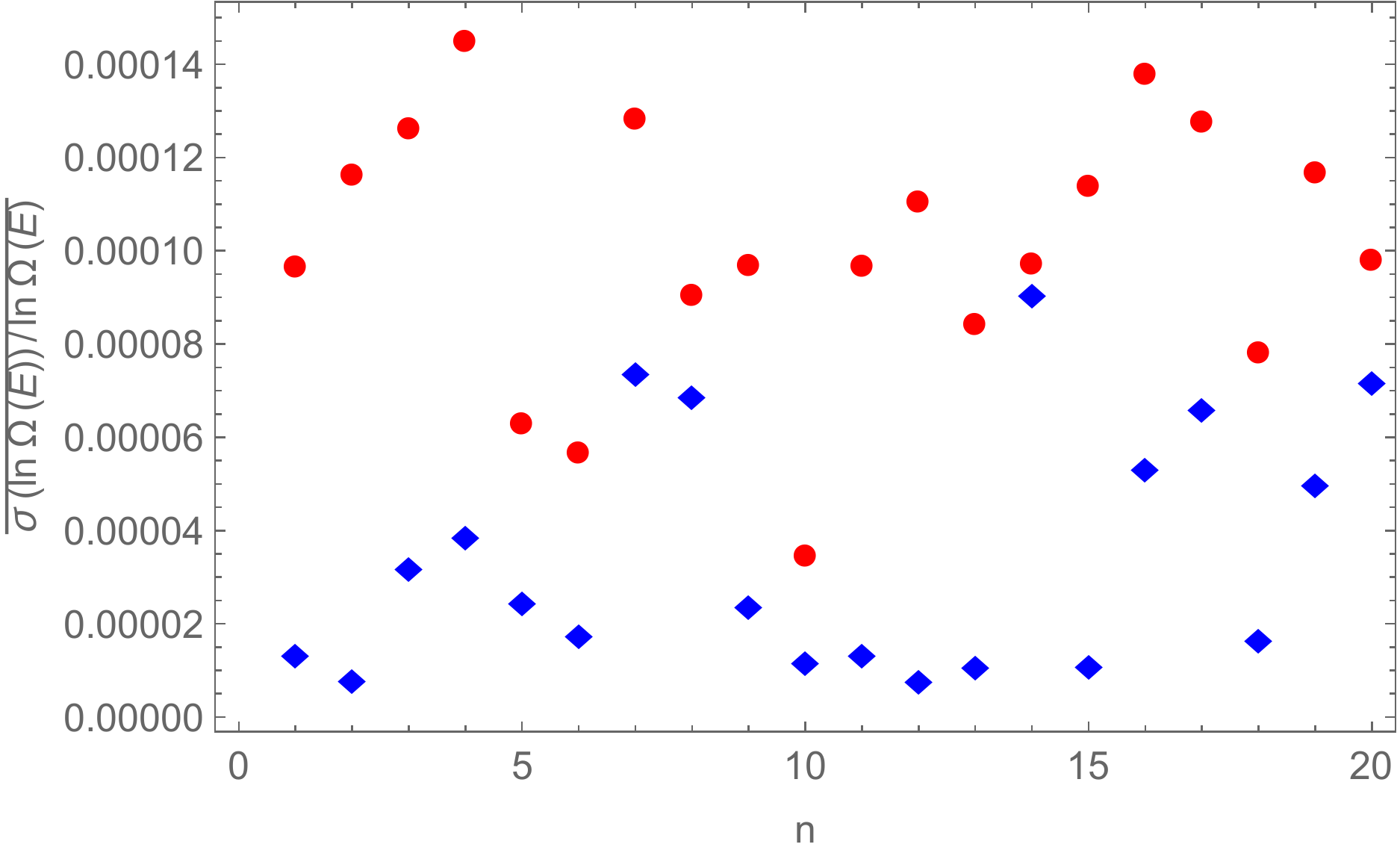}
\end{center}
\caption{Average relative standard deviations of level entropies,
  $\overline{\sigma[\ln(\Omega(E))]/\ln \Omega(E)}$, resulting from WL (red dots) and
  EPA (blue diamonds) simulations for $20$ 3D $\pm J$ Ising spin-glass samples of
  sizes $L=8$ (top) and $L=14$ (bottom), respectively. Here, the average is over
  energy levels. Note that for the WL result, we only included the 170 out of 200
  runs where the first phase completed successfully for all samples.}
\label{fig:3dsamples}
\end{figure}

As the samples considered are neither planar nor planted, we do not have access to
the exact DOS, and we hence quantify the success of the two algorithms in estimating
the DOS by determining the level of fluctuation in the estimates of $\ln \Omega(E)$
between independent runs, both for the WL-$1/t$ and EPA methods.  Specifically, we
estimated the relative standard deviation $\sigma[\ln \Omega(E)]/\ln \Omega(E)$ from
200 independent runs for each disorder sample.  In Fig.~\ref{fig:3dsamples} we show
this quantity, averaged over all energy levels, for 20 $\pm J$ 3D spin-glass samples
of the two system sizes considered. While for $L=8$ the WL runs yield slightly
smaller error bars, for $L=14$ the situation is reversed, with EPA resulting in 5
times smaller error bars, on average, corresponding to saving a factor of 25 in
run-time.

\subsection{Entropic sampling of problems of varying hardness}
\label{sec:varying_hardness}

Having ascertained that the EPA method can yield significantly better approximations
to the DOS of some hard problems in a given number of steps than the WL approach, it
is interesting to analyze more closely the actual distribution of performances of
these algorithms over the space of disorder realizations. The results above in
Figs.~\ref{fig:chimeraPAvsWL} and \ref{fig:toroidalPAvsWL} indicate the presence of
larger fluctuations in the quality of approximation for WL-$1/t$ as compared to EPA
across disorder realizations. To study this effect quantitatively, we used a set of
disorder samples classified according to their {\em thermal hardness\/}. A well
established measure of such hardness is the exponential autocorrelation or relaxation
time in parallel tempering simulations~\cite{hukushima:96a,sokal:97}. As very long
simulations are required to determine these time scales accurately, a number of proxy
quantities such as the so-called ``tunneling time'' are frequently used in practical
applications~\cite{katzgraber:06,bittner:08}. Here, we rely on a method developed in
the context of spin-glass simulations that analyses the dynamics of the random walk
of replicas in temperature space~\cite{banos:10} and extracts the corresponding
relaxation times $\tau$.

To benchmark the EPA algorithm against WL-$1/t$, we generated about $10^6$ random
instances on an $N=512$-spin Chimera graph (of which only 476 spins were used) and
measured the relaxation times $\tau$ of each instance with parallel
tempering\footnote{Specifically, we chose a temperature grid of the PT simulations
  consisting of $N_T=30$ temperatures. Temperatures with indices $i=1,2,\ldots,12$
  were uniformly distributed in the range $T_\mathrm{min}=0.045 \leq T_i \leq 0.2$,
  while the temperatures $T_i$ with $i=13,14,\ldots, N_T$ were spread evenly in the
  range $0.21 \leq T_i \leq T_\mathrm{max}=1.632$~\cite{scirep15:mm-hen}.}. Next, we
grouped together instances with similar classical hardness, i.e., similar relaxation
times $\tau$, $10^k \leq \tau \leq 3 \cdot 10^k$ for $k=3$, $4$, $5$, $6$ and
$7$. For each such `generation' of $\tau$, we randomly picked $100$ representative
instances for the benchmarking of the algorithm (only 14 instances with $k=7$ were
found). We then performed WL-$1/t$ simulations with a total of $10^{12}$ spin-flip
attempts for all samples, evenly split between one simulation each restricted to the
energy windows $[-900,-500]$ and $[-550,50]$, respectively (the ground-state energy
for these samples is roughly $E_0 \approx -800$). A pre-run of $2\times 10^{11}$
spin-flip attempts was again used to discover the range of possible energies for each
sample. All runs completed the first phase of the simulation here, owing to the use
of two energy windows. For the PA runs we used $R=2.1\times 10^6$, $\theta=10$,
$\alpha^\ast = 0.86$, corresponding to $N_\beta \approx 100$ temperature steps
($\beta_\mathrm{f} = 5$) and $10^{12}$ spin flip attempts. The DOS estimates from
both methods are only considered for $E \le 0$ and $\Omega(E) = \Omega(-E)$ is used
for $E > 0$. The resulting DOS estimate is normalized using the known total number of
states $2^N$.

\begin{figure}[!tb] 
\begin{center}
\includegraphics[width=0.7\textwidth]{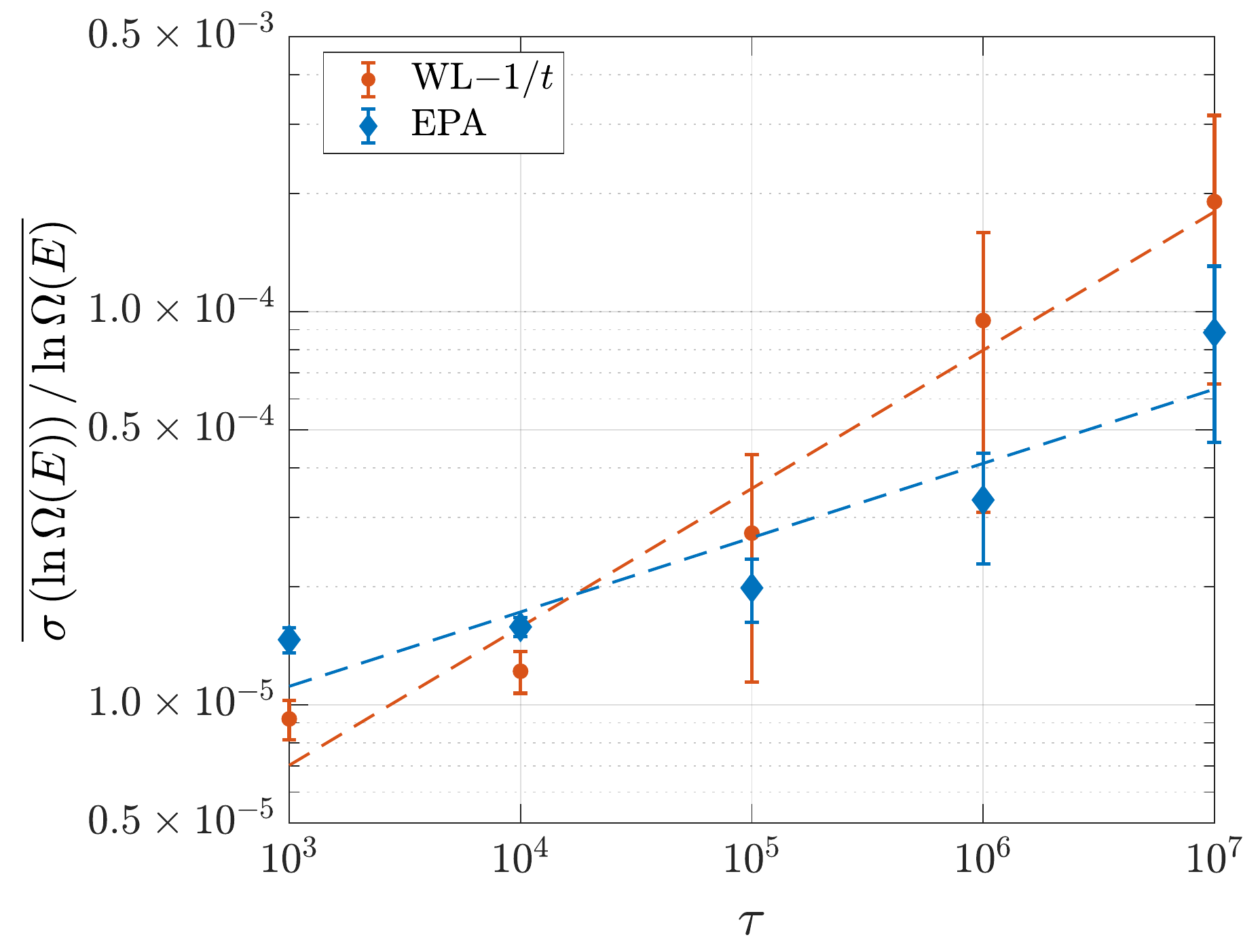}
\end{center}
\caption{\label{fig:isingHardness} Performance of the EPA and WL-$1/t$ estimates of
  the DOS on spin-glass samples of varying hardness. The plot shows the average standard
  deviation of level entropies as a function of the hardness group in terms of the
  parallel-tempering relaxation time $\tau$. WL-$1/t$ was run in each of the two
  energy windows, and the obtained DOS was normalized to the total number of states.
  The performance of WL-$1/t$ decreases sub-linearly with the problem difficulty,
  $\overline{\sigma(\ln\Omega(E))/\ln\Omega(E)} \sim \tau^{0.35}$, 
  while for EPA $\overline{\sigma(\ln\Omega(E))/\ln\Omega(E)} \sim \tau^{0.19}$.
  The results are averaged over 14 samples from each hardness class with the error
  bars resulting from the sample-to-sample fluctuations.  }
\end{figure}

As for the 3D samples, we compared the two algorithms by considering the relative
standard deviations $\sigma[\ln \Omega(E)]/\ln \Omega(E)$, averaged over all
energies. The resulting estimates are shown in Fig.~\ref{fig:isingHardness} for the
samples of the different hardness classes $k = 3$, $\ldots, 7$.  It is clear that
EPA is less affected by sample hardness than WL-$1/t$, with the growth in fluctuation
with sample hardness being much steeper for WL-$1/t$ than for EPA. Note that this
quantity only covers the effect of statistical errors, whereas the data in
Figs.~\ref{fig:chimeraPAvsWL} and \ref{fig:toroidalPAvsWL} considered the total
deviation from the exact results that also includes bias effects. Note also that the
sample-to-sample fluctuations, represented in the error bars of the data points in
Fig.~\ref{fig:isingHardness}, are significantly larger for WL-$1/t$ than for EPA. We
find that WL-$1/t$ and EPA have rather different behavior in sampling the DOS in
different energy ranges, with WL-$1/t$ being more focused on higher energies, for
details see~\ref{sec:energy-dependence}.

\begin{figure}[!tb] 
\begin{center}
  \includegraphics[width=0.7\textwidth]{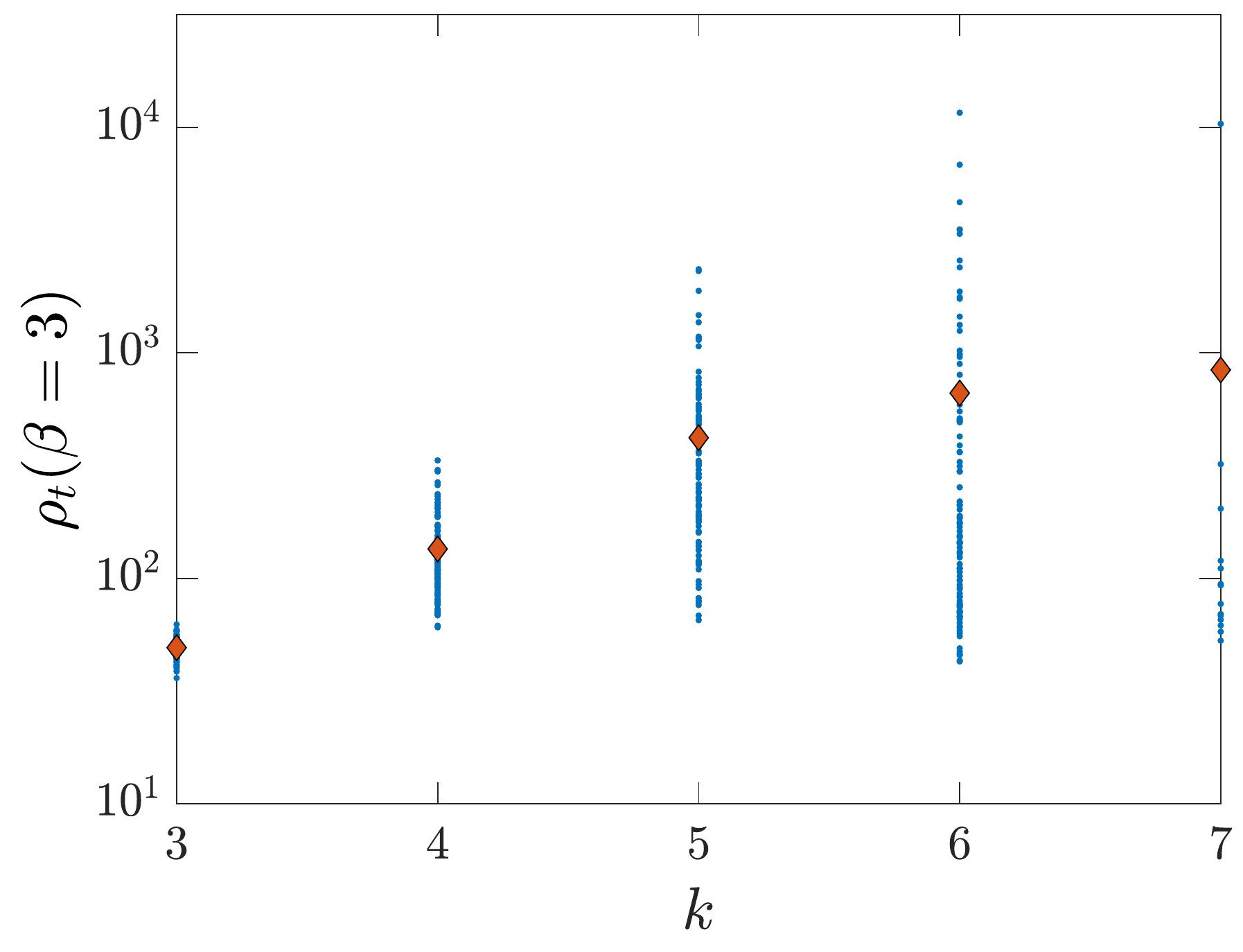}
\end{center}
\caption{\label{fig:rho} Mean-square family size $\rho_t$ at the lowest temperature
  in EPA for the varying-hardness samples (100 for $k=3$, $4$, $5$ and $6$ and 14 for
  $k=7$) for each hardness group. Larger average family sizes result from stronger
  correlations in the PA ensembles of replicas and hence indicate the difficulty of
  the algorithm in equilibrating the system.  The red diamonds mark the disorder
  average of $\rho_t$ at $\beta=3$ for each of the hardness groups.  }
\end{figure}

While in the present demonstration we used relaxation times from parallel tempering
for classifying sample hardness, it is worthwhile to note that EPA can itself provide
a hardness measure and thereby differentiate easy and hard samples. A few such
quantities have been previously proposed for population annealing~\cite{wang:15a}. We
consider here in particular the (temperature dependent) mean-square family size
$\rho_t$, defined as~\cite{wang:15a}
\begin{equation}
  \rho_t = R \sum_i{\frak n}_i^2,
  \label{eq:rhot}
\end{equation}
where ${\frak n}_i$ is the fraction of the current population that descends from the
$i$-th member of the initial population at $\beta_0 = 0$, while $R$ corresponds to
the initial population size. The quantity $R/\rho_t$ can be understood as an
effective population size, corresponding to the number of statistically independent
replicas, such that $R/\rho_t = R$ corresponds to a perfectly uncorrelated
population, while $R/\rho_t \to 0$ for the strongest correlations. These two limits
hence represent the easiest and hardest samples, for which one would expect
$\tau \to 0$ and $\tau \to \infty$, respectively, for parallel tempering.  A related
quantity that also takes the decorrelating effect of spin flips into account is the
effective population size $R_\mathrm{eff}$ defined in Ref.~\cite{weigel:17}. In
Fig.~\ref{fig:rho} we show a scatter plot of $\rho_t$ for 100 samples of each of the hardness
classes $k < 7$ and 14 samples for $k=7$, respectively. The disorder average of
$\rho_t$ at $\beta = 3$ is found to be $49$, $135$, $420$, $663$ and $840$ for $k=3$,
$4$, $5$, $6$, and $7$, respectively, indicating that while for the main part of the
distribution the hardnesses in EPA and parallel tempering are strongly correlated,
for the tails of the distribution the hardness in EPA increases more gently than that
found in parallel tempering.  As is demonstrated elsewhere, these intrinsic hardness
measures can be used to make population annealing simulations adaptive to the sample
hardness~\cite{amey:18,weigel:17a}. We note that the planted samples of
Sec.~\ref{sec:planted} have an average $\rho_t$ of $\approx 2000$ 
(see~\ref{appendix:rhot_planted}), indicating that planted samples 
of this type are much harder than random ones.

\section{Summary and discussion}

We have investigated the performance of sampling methods for estimating the DOS
for systems with complex free-energy landscapes, focusing on spin glasses
as the hardest problems among spin systems. We proposed a novel sampling technique
based on sequential Monte Carlo on a large population of copies and demonstrated that
it outperforms the most widely used entropic sampler, the Wang-Landau algorithm, in
the vast majority of cases. More importantly, the new approach shows better scaling
as the hardness of problems is increased (for example through considering larger
systems).

A notorious problem with benchmarking algorithms for estimating the DOS
lies in safely assessing convergence. Here we address this issue by considering
problems that are either of planar topology, in which case Pfaffian methods can be
used to determine the DOS exactly for systems with more than 1000
spins, or which have planted solutions such that the exact degeneracies of the ground
and first excited states can be calculated using a `bucket' algorithm. Both classes
provide hard optimization problems, implying a very non-trivial benchmark. In
addition, we also considered more general problems such as stochastic spin-glass
samples on Chimera graphs sorted by thermal hardness as well as the most challenging
case of three-dimensional spin-glass instances of up to $14\times 14\times 14$ spins.

One essential advantage of the approach based on population annealing is that it
does not require any prior knowledge about the energy spectrum, which in contrast
needs to be acquired in an additional pre-run for the Wang-Landau
method. Furthermore, the well-known and delicate problem of dividing the energy range
into windows that is the only way of making Wang-Landau simulations converge for the
more challenging cases, is completely absent for entropic population annealing. The
difficulty of premature and false convergence that plagues Wang-Landau and related
methods is not so much of an issue for the newly introduced technique, where a
re-distribution of weights can occur at all stages of the algorithm. In fact, for
the EPA method it is easily possible to monitor equilibration from intrinsic
properties and only output the DOS for energies where thermalization could be
ensured. The main advantage of the approach, however, lies in the ideal suitability
for massively parallel calculations, where given sufficient parallel resources the
accuracy of the approximation can be arbitrarily improved at a constant wall-clock
time by increasing the size of the population or combining the outcome of independent
runs in a weighted average.

The specific spin system we study, namely spin glasses, is very relevant as current
experimental quantum annealers attempt to solve precisely this type of problem. With
questions still lingering about which distribution these devices sample from
\cite{marshallrieffelhen:2017}, it is important to have an accurate tool to estimate
the DOS (for instance to understand
thermalization~\cite{marshallrieffelhen:2017}). For this problem we specifically
consider instances with vastly different hardnesses, confirming that the accuracy of
the technique proposed degrades significantly less for harder samples that previous
approaches. We note that the entropic population annealing approach is in no way
specific to the spin systems considered here, and it can be straightforwardly
generalized to other problems such as lattice polymers and, with the help of binning
or spectral methods \cite{farris:19}, to cases with continuous degrees of freedom.

Our results are very promising as we could clearly show that i) for a range of
problems with complex energy landscapes existing sampling methods for the DOS
are difficult to set up and do not converge very reliably, especially for hard
samples; ii) a dependable method for the benchmarking of entropic samplers on spin
glasses is now available, which we hope will drive research forward in finding even
better algorithms; and iii) the entropic population annealing (EPA) algorithm devised
here allows for the reliable sampling of large-scale frustrated systems. We therefore
trust that the algorithm will become a useful tool for DOS calculation
in condensed matter physics, quantum computing and other areas of research.

\ack

Part of the computing resources were provided by the USC Center for High Performance
Computing and Communications and the Oak Ridge Leadership Computing Facility at the
Oak Ridge National Laboratory, which is supported by the Office of Science of the
U.S.  Department of Energy under Contract No. DE-AC05-00OR22725.  The research is
based upon work (partially) supported by the Office of the Director of National
Intelligence (ODNI), Intelligence Advanced Research Projects Activity (IARPA), via
the U.S. Army Research Office contract W911NF-17-C-0050. The views and conclusions
contained herein are those of the authors and should not be interpreted as
necessarily representing the official policies or endorsements, either expressed or
implied, of the ODNI, IARPA, or the U.S. Government. This material is based on
research sponsored by AFRL under agreement number FA8750-18-1-0109. The
U.S. Government is authorized to reproduce and distribute reprints for Governmental
purposes notwithstanding any copyright annotation thereon.  LB was supported by grant
No.\ 14-21-00158 from the Russian Science Foundation. LB and MW acknowledge support
by the European Commission through the IRSES network DIONICOS under Contract No.\
PIRSES-GA-2013-612707.

\bibliography{paper}

\appendix

\section{The DOS estimator}
\label{sec:dos-estimator}

In this section we outline the derivation of the estimator \eqref{eq:MHReq1} for the
DOS. We initially follow the reasoning of
Ref.~\cite{ferrenberg:89a}. In population annealing, the histogram of energies at
temperature $\beta_i$, $\hat{H}_{\beta_i}(E)$ or in short $\hat{H}_i$, is an
estimator of the equilibrium probability density of internal energies,
\begin{equation}
  \left\langle \frac{\hat{H}_i(E)}{R_i} \right\rangle = \frac{1}{Z_i}\Omega(E)
  e^{-\beta_i E}.
  \label{eq:energy-distribution}
\end{equation}
Hence an estimate of the DOS from a single histogram is given by
\begin{equation}
  \hat{\Omega}_i(E) = \frac{\hat{H}_i(E)}{R_i} \frac{Z_i}{e^{-\beta_i E}}.
  \label{eq:omega-and-histogram}
\end{equation}
To make use of the histograms at different temperature steps, we take a weighted
average,
\begin{equation}
  \hat{\Omega}(E) = \sum_i \alpha_i \hat{\Omega}_i(E).
  \label{eq:weighted-average}
\end{equation}
As a simple calculation shows \cite{brandt:book}, for independent individual
estimates a minimum variance of the result is achieved when choosing the weights
\[
  \alpha_i = \frac{1/\sigma^2[\hat{\Omega}_i(E)]}{\sum_j 1/\sigma^2[\hat{\Omega}_j(E)]}.
\]
From Eq.~\eqref{eq:omega-and-histogram} we deduce that
\begin{equation}
  \sigma^2[\hat{\Omega}_i(E)] = \frac{Z_i^2}{R_i^2}e^{2\beta_i E}
  \sigma^2[\hat{H}_i(E)],
  \label{eq:relation-between-variances}
\end{equation}
and hence the variance-optimized weighted average \eqref{eq:weighted-average} becomes
\begin{equation}
  \label{eq:dos-weighted-average}
  \hat{\Omega}(E) = \frac{\sum_i R_i Z_i^{-1} e^{-\beta_i
      E}\hat{H}_i(E)\sigma^{-2}[\hat{H}_i(E)]}{\sum_j R_j^2 Z_j^{-2} e^{-2\beta_j
      E} \sigma^{-2}[\hat{H}_j(E)]}.
\end{equation}
Noting that $Z_i = \exp[-\beta_i F(\beta_i)]$ and that
$\sum_E \hat{\Omega}(E) \exp[-\beta E]$ is an estimator of $Z = \exp[-\beta F]$, one
can write the following equation,
\begin{equation}
  e^{-\beta_k \hat{\cal F}_k} = \sum_E \frac{\sum_i R_i \exp[\beta_i\hat{\cal F}_i-\beta_i
      E]\hat{H}_i(E)\sigma^{-2}[\hat{H}_i(E)]}{\sum_j R_j^2 \exp[2\beta_j\hat{\cal F}_j-2\beta_j
      E] \sigma^{-2}[\hat{H}_j(E)]} e^{-\beta_k E},
  \label{eq:self-consistency}
\end{equation}
which can be read as a set of $N_\beta$ self-consistency conditions for the
parameters $\hat{\cal F}_k$ that represent the free energies at the inverse
temperatures $\beta_k$. Eq.~\eqref{eq:self-consistency} can be solved iteratively by
evaluating the right hand side for each $k$ to receive updated values of $\hat{\cal
  F}_k$. Convergence is very slow (and might even fail) if starting with initial
values for $\hat{\cal F}_k$ that are very far from the solution. If one starts with
$\hat{\cal F}_k = \hat{F}(\beta_k)$ according to Eq.~\eqref{eq:free-energy}, however,
which already provides very accurate estimates of the free energies, convergence is
typically achieved in less than ten iterations.

It remains to discuss how to determine the variances
$\sigma^2[\hat{H}_i(E)]$. Without further assumptions, these can be estimated via a
jackknife analysis \cite{efron:book}, i.e., by dividing the populations at all
temperature steps into jackknife blocks. Calculating $\hat{H}_i(E)$ for each
jackknife block then allows one to use the jackknife estimator of variance, energy by
energy\footnote{Note that in order to apply the jackknife procedure one should
  consider an observable with a finite expectation value in the limit of an infinite
  number of measurements, such as $\hat{H}_i(E)/R_i$.}. Simpler expressions can be
derived with further assumptions, as we will discuss now. If all members of the
population were independent of each other, the entries in a histogram would follow a
Poisson distribution and hence the variance was
\begin{equation}
  \label{eq:poisson-approximation}
  \sigma^2[\hat{H}_i(E)] = \langle \hat{H}_i(E) \rangle = R_i Z_i^{-1} \Omega(E)
  e^{-\beta_i E},
\end{equation}
where the second equality follows from
Eq.~\eqref{eq:energy-distribution}. Substituting this into
Eq.~\eqref{eq:dos-weighted-average} one finds
\begin{equation}
  \label{eq:dos-weighted-average2}
   \hat{\Omega}(E) =
   \frac{\sum_i \hat{H}_i(E)}{\sum_j
    R_j \exp[\beta_j \hat{\cal F}_j - \beta_j E]},
\end{equation}
which corresponds to the estimator \eqref{eq:MHReq1} introduced in the main text when
using the single-temperature estimate $\hat{\cal F}_j = \hat{F}(\beta_j)$. While this
estimate can be further improved, in principle, via the iterations
\eqref{eq:self-consistency}, we find that for the small temperature steps used in our
runs the improvement is quite small, and the simpler approach of
Eq.~\eqref{eq:dos-weighted-average2} with $\hat{\cal F}_j = \hat{F}(\beta_j)$ already
yields excellent results.

In reality different replicas of the population are not independent of each other as
the resampling introduces correlations. One might argue that as this reduces
the effective population size by a factor $R_{\mathrm{eff},i}/R_i$ \cite{weigel:17a},
Eq.~\eqref{eq:poisson-approximation} should be replaced by
\begin{equation}
  \label{eq:poisson-approximation2}
  \sigma^2[\hat{H}_i(E)] = \frac{R_i}{R_{\mathrm{eff},i}} \langle \hat{H}_i(E)
  \rangle = \frac{R_i^2}{R_{\mathrm{eff},i}} Z_i^{-1} \Omega(E)
  e^{-\beta_i E},
\end{equation}
such that Eq.~\eqref{eq:dos-weighted-average2} is replaced by
\begin{equation}
  \label{eq:dos-weighted-average3}
  \hat{\Omega}(E) =
  \frac{\sum_i \frac{R_i}{R_{\mathrm{eff},i}} \hat{H}_i(E)}{\sum_j
    \frac{R_j}{R_{\mathrm{eff},j}} R_j \exp[\beta_j {\cal F}_j - \beta_j E]}.
\end{equation}
In the absence of an estimate for $R_{\mathrm{eff},i}$, it is also possible to revert
to $R_i/{\rho_{t,i}}$ as defined in Eq.~\eqref{eq:rhot}, which in general provides a
lower bound for $R_{\mathrm{eff},i}$ \cite{weigel:17}. This approximation assumes,
however, that the only effect of correlations is to reduce the effective number of
events and, in particular, that this effect is independent of the energy level
$E$. 
Preliminary tests of applying this variant of the DOS estimator yielded mixed 
results with decreases in $\Delta$ of Eq.~\eqref{eq:delta} of
at most 10\%, so the effect appears to be rather weak.

Finally, we should also take into account the fact that the populations at different
temperature steps are correlated\footnote{This is in contrast to the usual situation
  in multi-histogram reweighting, where the individual histograms belong to
  independent simulations.}. In this case the average \eqref{eq:weighted-average}
should be performed with weights \cite{brandt:book,weigel:09,weigel:10}
\[
  \alpha_i = \frac{\sum_j \cov^{-1}(\hat{\Omega}_i, \hat{\Omega}_j)}
  {\sum_{ij} \cov^{-1}(\hat{\Omega}_i, \hat{\Omega}_j)}.
\]
Here,
\[
  \cov(\hat{\Omega}_i, \hat{\Omega}_j) = \langle \hat{\Omega}_i(E)
  \hat{\Omega}_j(E)\rangle - \langle \hat{\Omega}_i(E) \rangle \langle
  \hat{\Omega}_j(E)\rangle
\]
denotes the covariance matrix of the DOS estimates $\hat{\Omega}_i(E)$ from different
temperature steps, and $\cov^{-1}$ is the corresponding inverse matrix. We note that,
in generalization of Eq.~\eqref{eq:relation-between-variances}, one has
\[
  \cov[\hat{\Omega}_i(E), \hat{\Omega}_j(E)] = \frac{Z_i Z_j}{R_i R_j}
  e^{(\beta_i+\beta_j)E} \cov[\hat{H}_i(E), \hat{H}_j(E)].
\]
As for the variances, the covariances of the energy histograms can be estimated from
a jackknife analysis over the populations \cite{efron:book,weigel:10}. The
corresponding expression for the optimized DOS estimate is then generalized to
\begin{equation}
  \label{eq:dos-weighted-average4}
  \hat{\Omega}(E) = \frac{\sum_{i,j} R_j Z_j^{-1} e^{-\beta_j
      E}\hat{H}_i(E) \cov^{-1}[\hat{H}_i(E), \hat{H}_j(E)]}{\sum_{i,j} R_i R_j
    Z_i^{-1} Z_j^{-1} e^{-(\beta_i+\beta_j) E} \cov^{-1}[\hat{H}_i(E), \hat{H}_j(E)]}.
\end{equation}

\section{Estimating statistical errors and biases}
\label{sec:error-appendix}

\begin{figure}[tb!]
\begin{center}
  \includegraphics[width=0.7\textwidth]{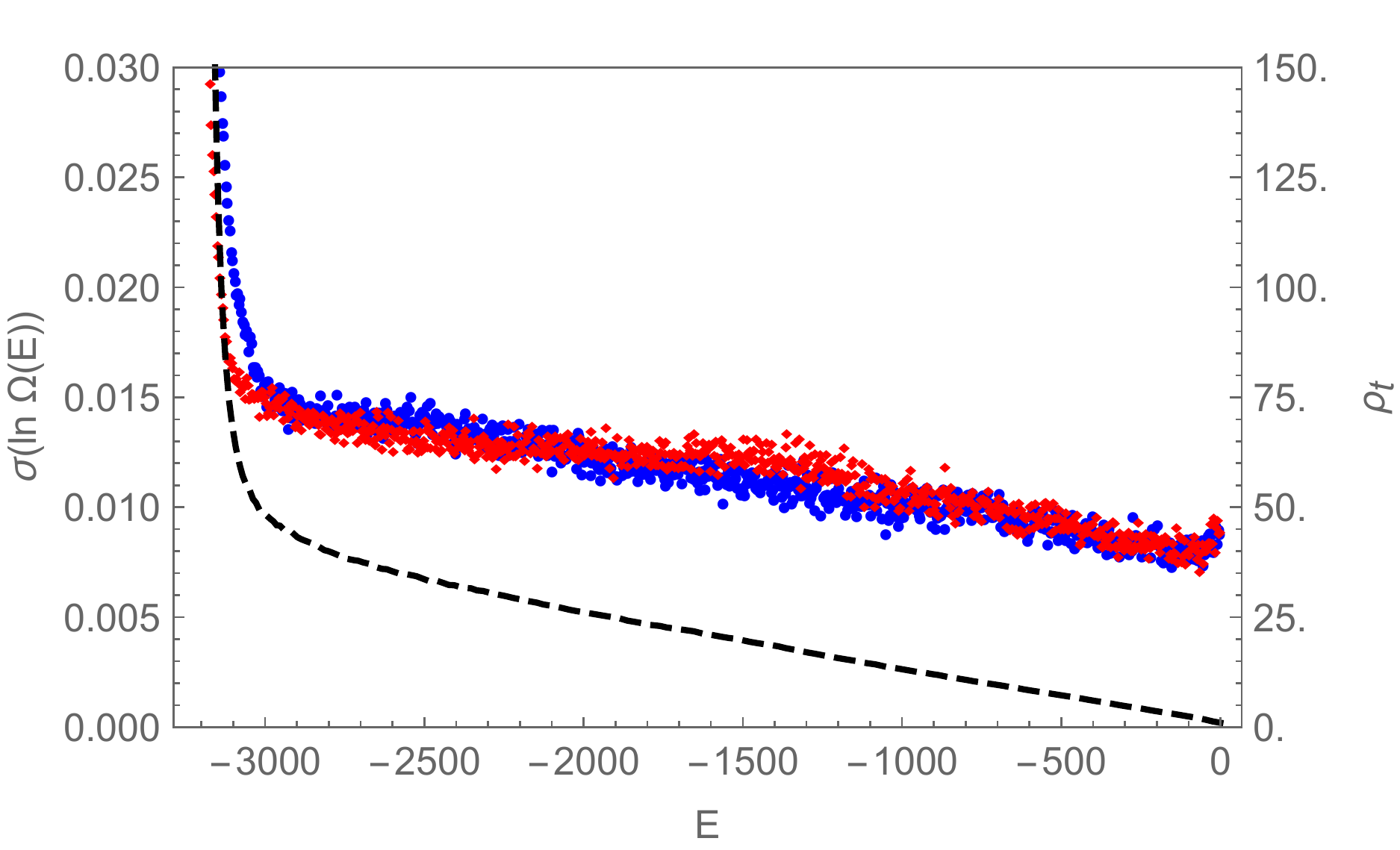}
\end{center}
\caption{Statistical error of the logarithm of the DOS estimate for an $L=48$
  toroidal sample from EPA with $R=100\,000$, $\theta=10$ estimated from a
  jackknife analysis (red diamonds) as compared to the reference estimate from 200
  independent runs (blue dots). Both estimates are in excellent agreement down to
  energies where the simulation starts to fall out of equilibrium as indicated by the
  steep increase of $\rho_t$ (dashed line, right scale).}
\label{fig:jackknife}
\end{figure}

Statistical errors in the DOS estimator $\hat{\Omega}(E)$ provided in
Eq.~\eqref{eq:MHReq1} or the variants discussed in \ref{sec:dos-estimator} can be
estimated by considering the statistics of different runs. As the correct way of
combining different runs is through extending the summations on the right hand side of
Eq.~\eqref{eq:MHReq1} over all the temperature points of all runs, combined with the
corresponding free-energy estimates provided by Eq.~\eqref{eq:free-energy} instead of
taking a plain average over the estimates $\hat{\Omega}(E)$ of individual runs, it is
important to estimate errors from a jackknife or bootstrap analysis over the runs
instead of the standard estimator of the sample mean in order to minimize bias.

\setcounter{footnote}{0}

Alternatively, it it possible to estimate statistical errors from a single run of a
sufficiently large population following the arguments of Ref.~\cite{weigel:17}. This
amounts to a simultaneous jackknife analysis over the populations at all
temperatures. As discussed in detail in Ref.~\cite{weigel:17}, such an analysis can
be justified if a linear order of the members of the population is assumed, and
off-spring configurations in resampling are placed next to each other in the array
of replicas. As one then generally observes an exponential decay of correlations with
the index distance in replica space, sufficiently large blocks are statistically
effectively independent of each other, and the jackknife estimator for the variance
of the mean can be applied. For the case of the DOS estimator of
Eq.~\eqref{eq:MHReq1} this implies that one divides the (linearly ordered)
populations at all temperature steps into $n$ blocks ($n=100$ is often a good choice)
and then applies Eqs.~\eqref{eq:free-energy} and \eqref{eq:MHReq1} to all data apart
from the replicas in block $s=1$, $\ldots$, $n$ to arrive at estimates
$\ln \hat{\Omega}_{(s)}(E)$\footnote{As $\Omega(E)$ spans many orders of magnitude,
  it is normally much more reasonable to consider $\ln \Omega(E)$, and we hence here
  also estimate the error bars of this latter quantity.}. The variance of the mean
(squared error bar) is then estimated by \cite{efron:book}
\begin{equation}
  \hat{\sigma}^2[\ln\hat{\Omega}(E)] = \frac{n}{n-1} \sum_{i=1}^n
  \left[\ln\hat{\Omega}_{(s)}(E) - \ln\hat{\Omega}_{(\cdot)}(E)\right]^2,
  \label{eq:jackknife-error}
\end{equation}
where
\[
  \ln\hat{\Omega}_{(\cdot)}(E) = \frac{1}{n}\sum_{s=1}^n \ln\hat{\Omega}_{(s)}(E).
\]
As is illustrated in Fig.~\ref{fig:jackknife} this provides an accurate estimate of
statistical errors as long as the total simulation remains in thermal equilibrium.

Regarding bias effects (i.e., lack of thermalization), we note that EPA as a
sequential Monte Carlo method behaves differently to Markov chain samplers and the
closely related WL approach. Whereas in the latter the results of a simulation that
was not properly equilibrated will be affected by biases in its entirety, a
population annealing simulation of a frustrated system will fall out of equilibrium
at a certain threshold temperature, and consequently results derived from the
populations at and below this temperature will be affected by biases. Conversely,
however, results only incorporating data above this threshold are unaffected. For the
DOS estimators studied here a strategy to avoid bias hence consists of only including
histograms from the well-equilibrated regime in the estimate \eqref{eq:MHReq1}. Here,
equilibration can be ensured by monitoring heuristic criteria such as
$R_0 > 100\rho_t$ \cite{wang:15a,amey:18} that can be evaluated on the fly to
determine a stopping temperature.

\section{The square-lattice Ising ferromagnet}
\label{sec:ising}

\begin{figure}[tb!]
\begin{center}
  \includegraphics[width=0.7\textwidth]{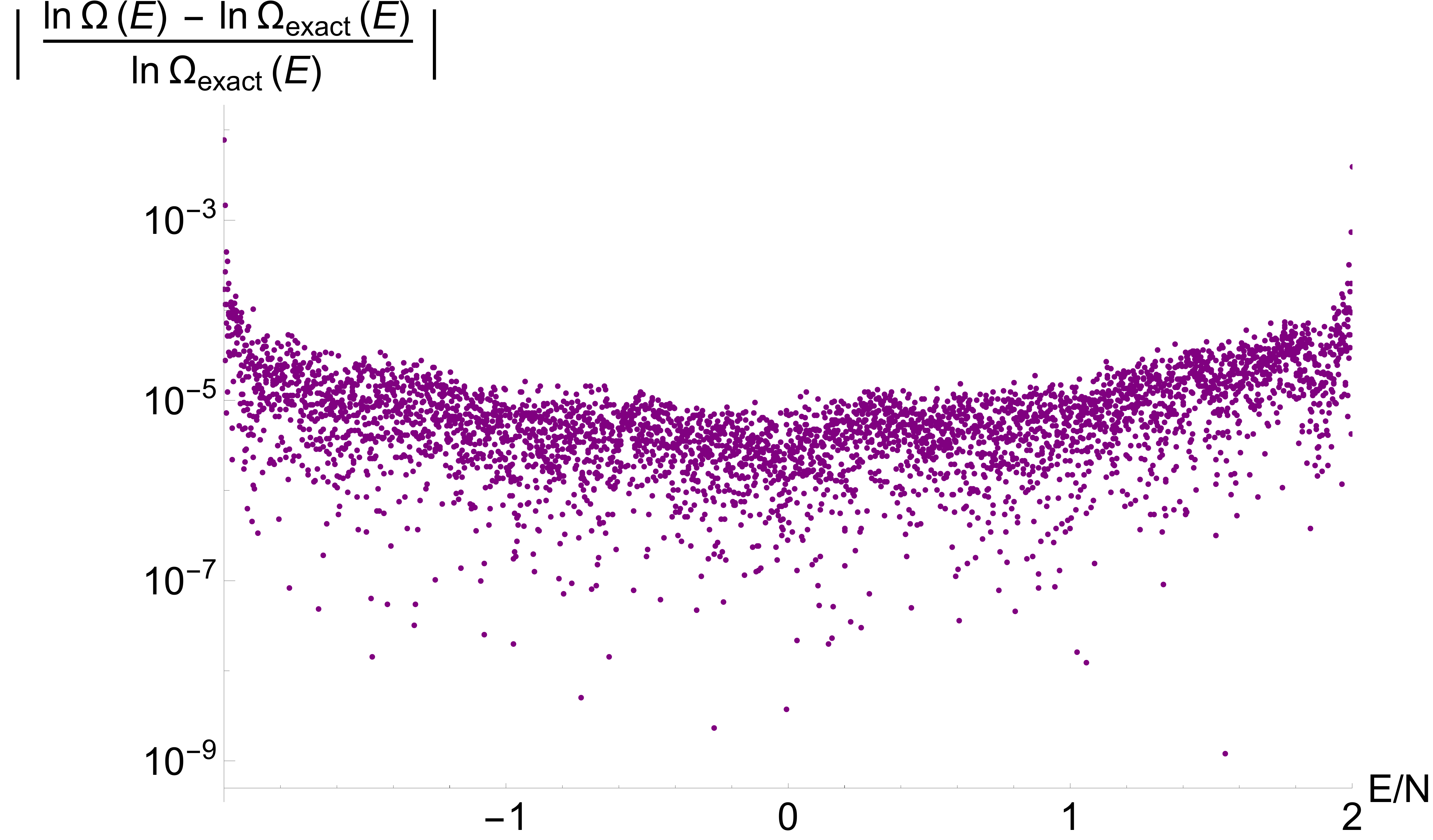}
\end{center}
  \caption{Relative deviation of level entropies from the exact result for simulating
    the  Ising ferromagnet on an $L=64$ square lattice using EPA with  $R=100\,000$, $\theta=10$, $\alpha^\ast=0.86$, corresponding
    to $N_\beta\approx 230$ temperature steps and a total of $9.4\times 10^{11}$ spin flips.
  }
  \label{fig:isingDOS}
\end{figure}

The case of the Ising ferromagnet on the square lattice, corresponding to the
Hamiltonian (\ref{eq:Hp}) with $J_{ij}=1$ for all nearest-neighbor pairs $(i,j)$, has
served as a standard benchmark for entropic samplers since they were first considered
\cite{wang:01a}. In this case, the DOS can be exactly computed using methods that are
somewhat simpler than those of Ref.~\cite{galluccio:00}, see
Ref.~\cite{beale:96a}. Fig.~\ref{fig:isingDOS} shows the relative deviation of
$\ln \Omega(E)$ obtained via EPA from the exact DOS for system size $L=64$. Similar
plots can be found for the WL method in Refs.~\cite{wang:01a,wang:01b} and
subsequently in many papers on improved methods. The parameters used for the EPA run
are $R=100\,000$, $\theta=10$, and $\alpha^\ast=0.86$, corresponding to
$N_\beta\approx 230$ temperature steps and $9.4\times 10^{11}$ spin flips. It is
apparent that a high accuracy is achieved across the whole energy range.

\begin{table}[tb!]
\begin{center}
\begin{tabular}{|c|c|c|}
  \hline
  method & $\overline{\Delta}$ & $\sigma(\overline{\Delta})$\\
  \hline
  WL-1/t & $2.67\times 10^{-5}$ & $1.55\times 10^{-5}$\\
  \hline
  EPA, $\theta=1$ & $8.55\times 10^{-5}$ & $6.80\times 10^{-5}$ \\
  EPA, $\theta=5$ & $2.13\times 10^{-5}$ & $1.29\times 10^{-5}$ \\
  EPA, $\theta=10$ & $2.66\times 10^{-5}$ & $1.69\times 10^{-5}$ \\
  EPA, $\theta=20$ & $3.53\times 10^{-5}$ & $1.82\times 10^{-5}$ \\
  EPA, $\theta=50$ & $5.43\times 10^{-5}$ & $2.85\times 10^{-5}$ \\
  \hline
\end{tabular}
\end{center}
\caption{Average deviation $\overline{\Delta}$ for the two-dimensional Ising model
  with $L=32$. The deviation is averaged over 200 independent runs of the EPA and WL
  algorithms. Each of the runs performed $N=9.82\times 10^{11}$ spin flips.  Also
  shown is the standard deviation of $\overline{\Delta}$.  }
\label{tab:isingDOS}
\end{table}

In Table~\ref{tab:isingDOS} we compare the average deviation $\Delta$ according to
Eq.~(\ref{eq:delta}) for different runs of the WL-$1/t$ and EPA algorithms.  The
WL-$1/t$ runs obtained the full DOS, while EPA simulations obtained the degeneracies
of energy levels with $E\lesssim 0$ and we exploited the symmetry
$\Omega(-E) = \Omega(E)$. The accuracy of WL-$1/t$ is approximately the same as the
accuracy of EPA for $5 < \theta < 10$, for the same number of spin flips.
Introducing an automatic adaptation of $\theta$ for EPA results in an additional
approximately threefold reduction of $\overline{\Delta}$~\cite{weigel:17a}.

\section{Normalization of the DOS estimates}
\label{sec:normalization}

\begin{figure*}[tb!]
\begin{center}
  \includegraphics[width=0.95\textwidth]{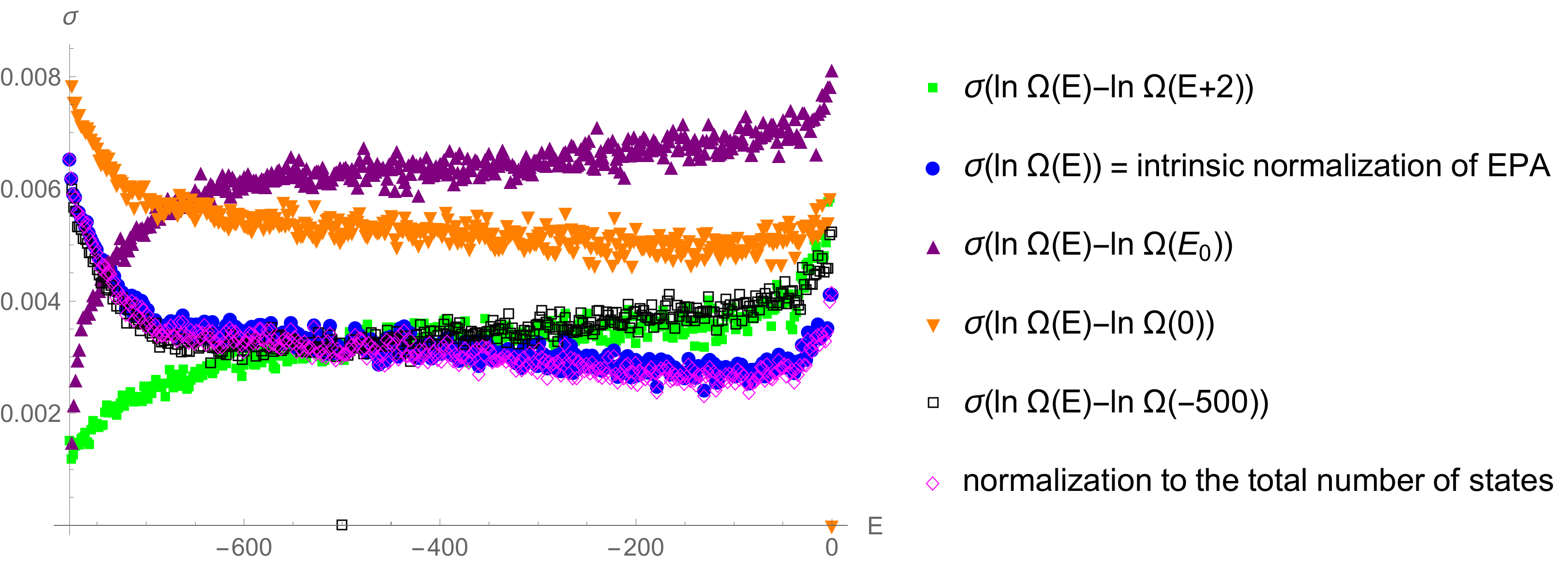}
  \includegraphics[width=0.95\textwidth]{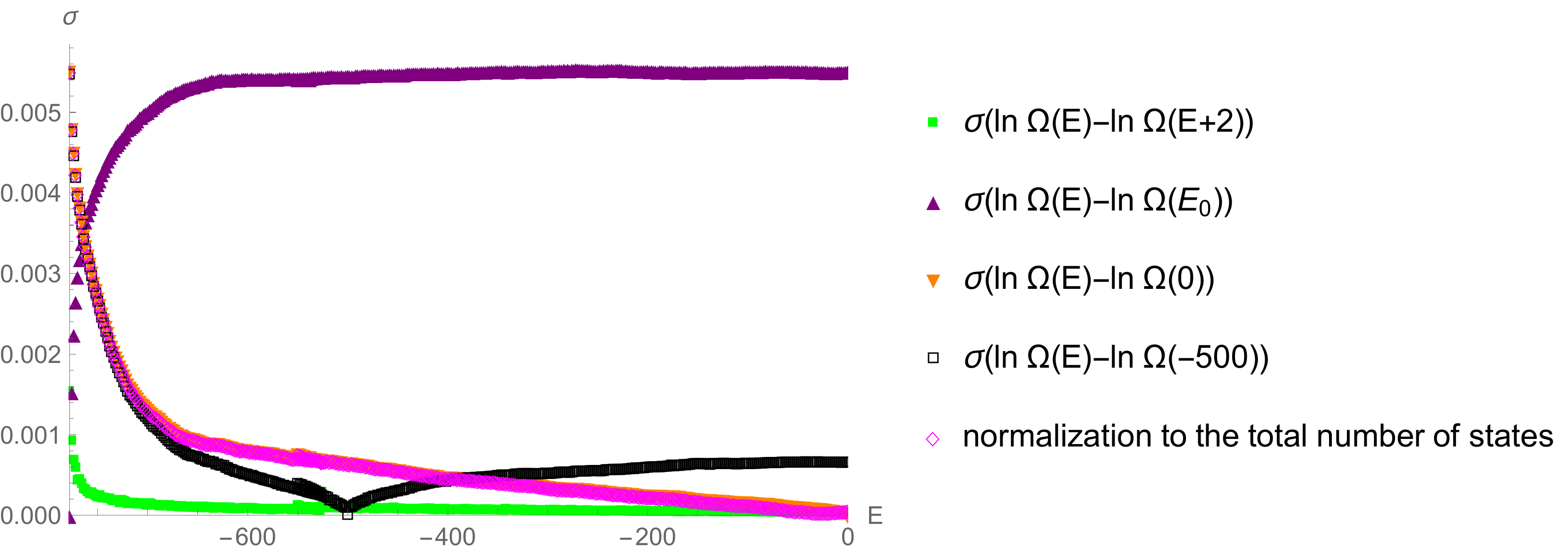}
\end{center}
  \caption{Top: Standard deviations of the estimated level entropies from EPA simulations
    using different normalization schemes. A single sample from the hardness class
    $k = 3$ is considered.  The parameters of the EPA simulation are described in
    the caption to Fig.~\ref{fig:C} below. Bottom: The analogous plot for WL-$1/t$. The run
    parameters are described in Sec.~\ref{sec:varying_hardness}.}
  \label{fig:D}
\end{figure*}

\begin{figure*}[tb!]
\renewcommand{\thefigure}{E1}
\begin{center}
  \includegraphics[width=0.49\textwidth]{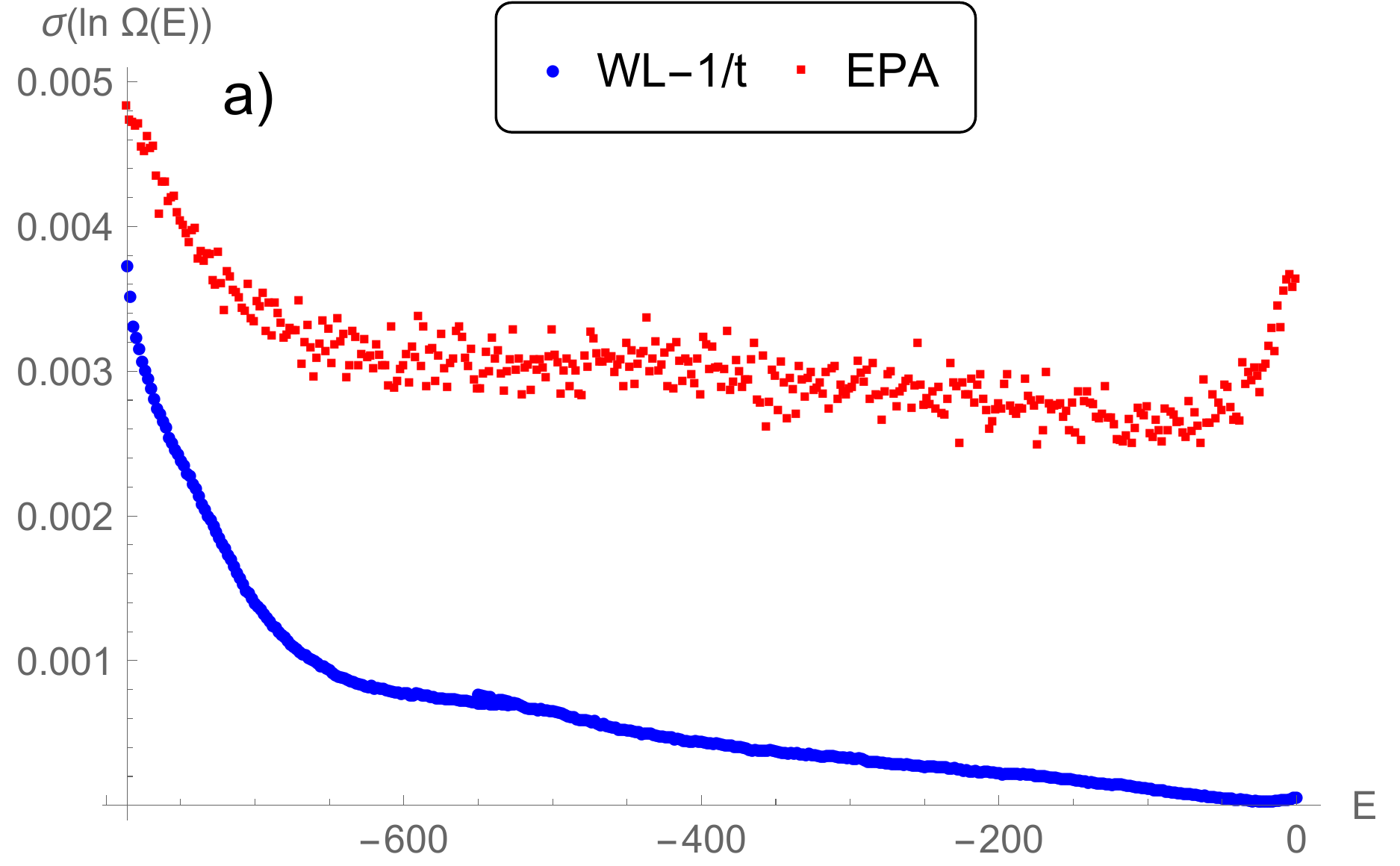}
  \includegraphics[width=0.49\textwidth]{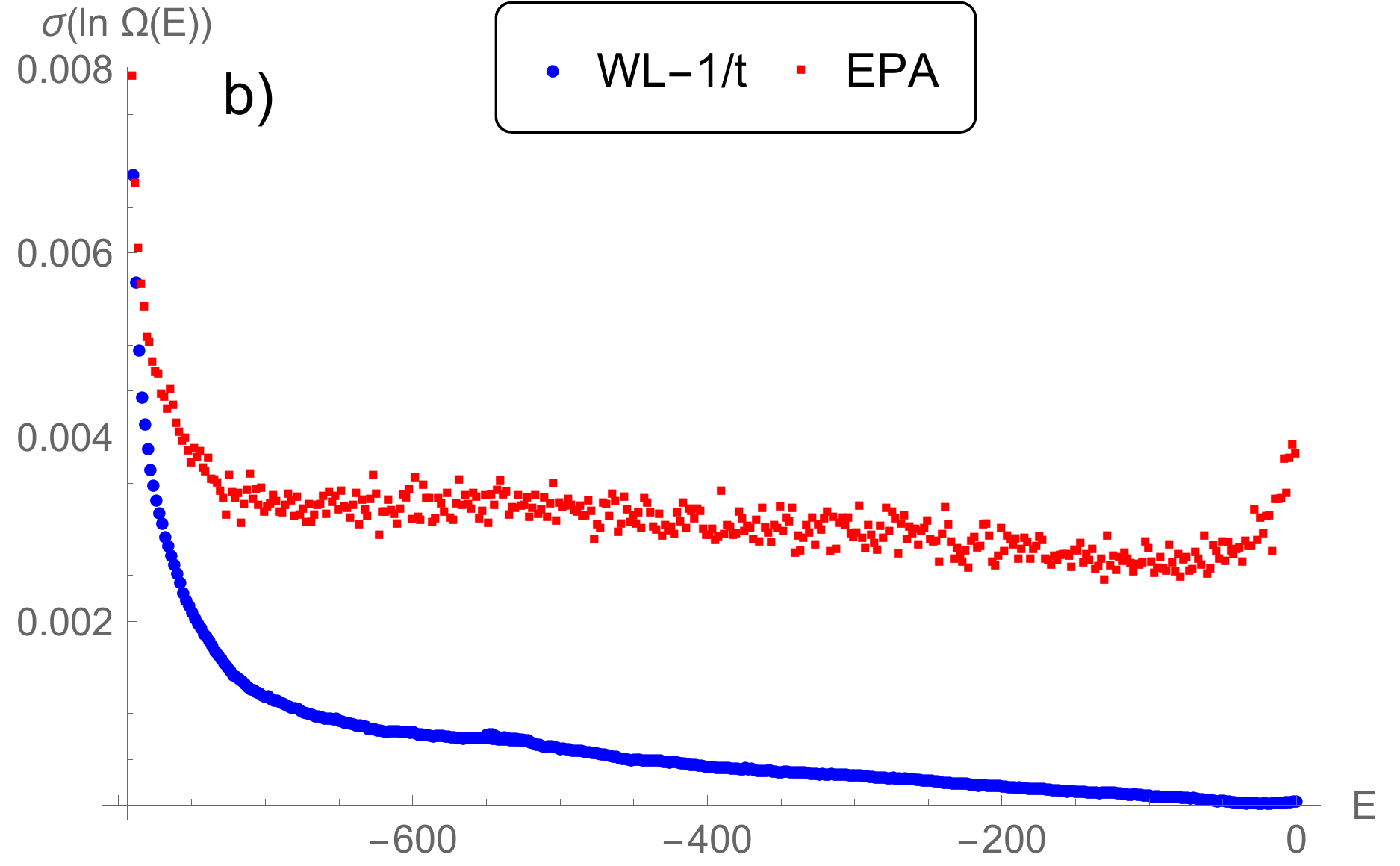}
  \includegraphics[width=0.49\textwidth]{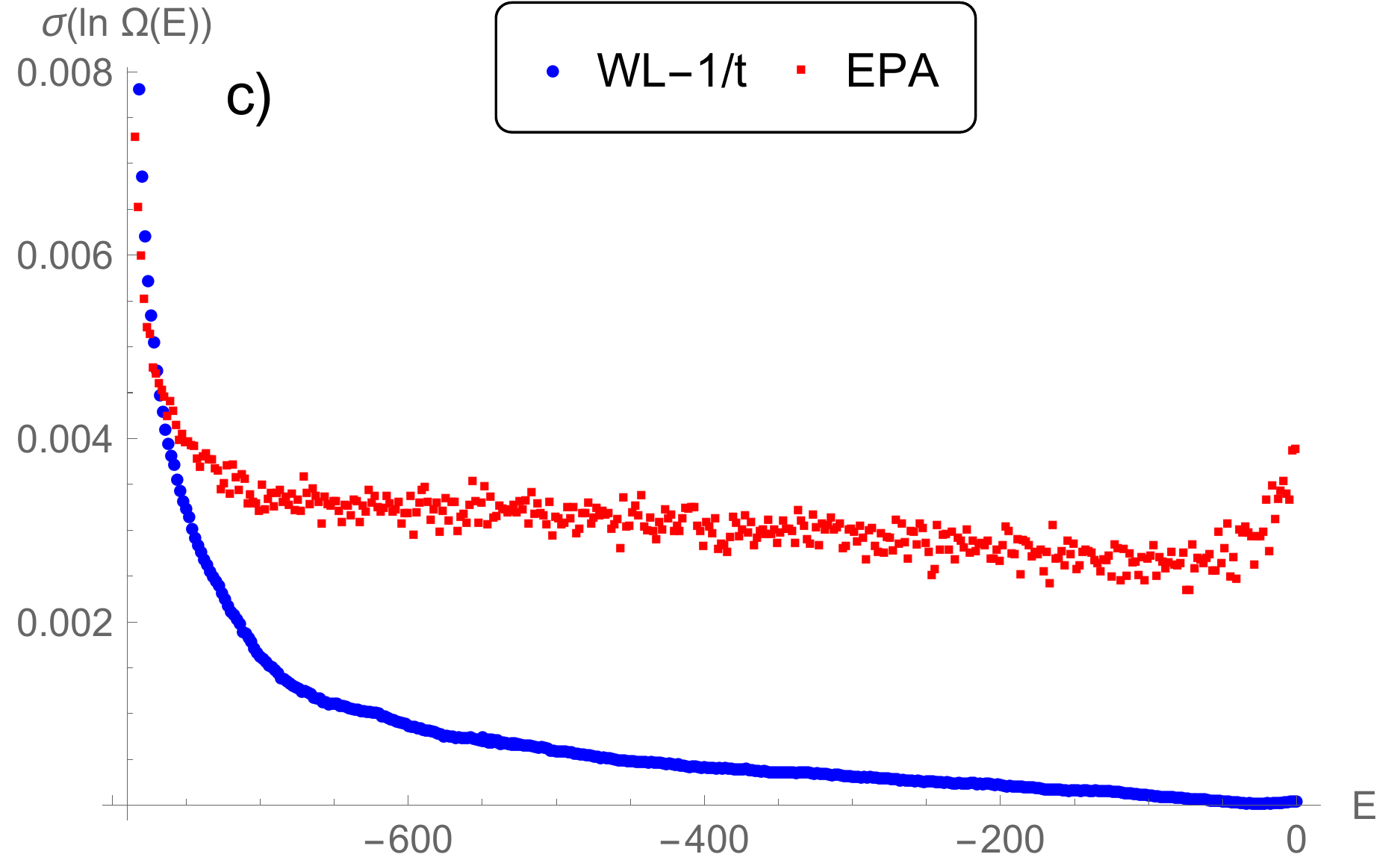}
  \includegraphics[width=0.49\textwidth]{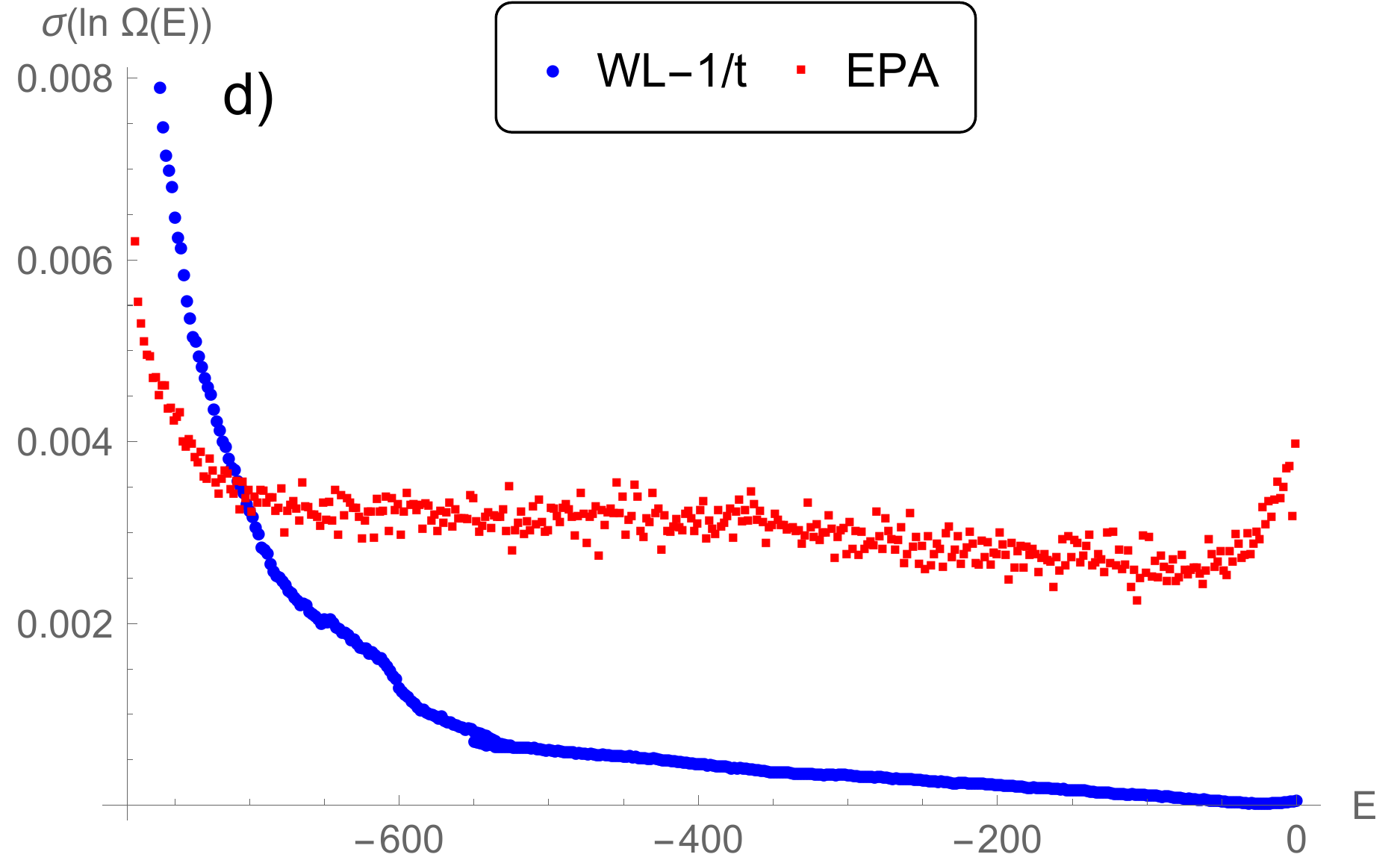}
  \includegraphics[width=0.49\textwidth]{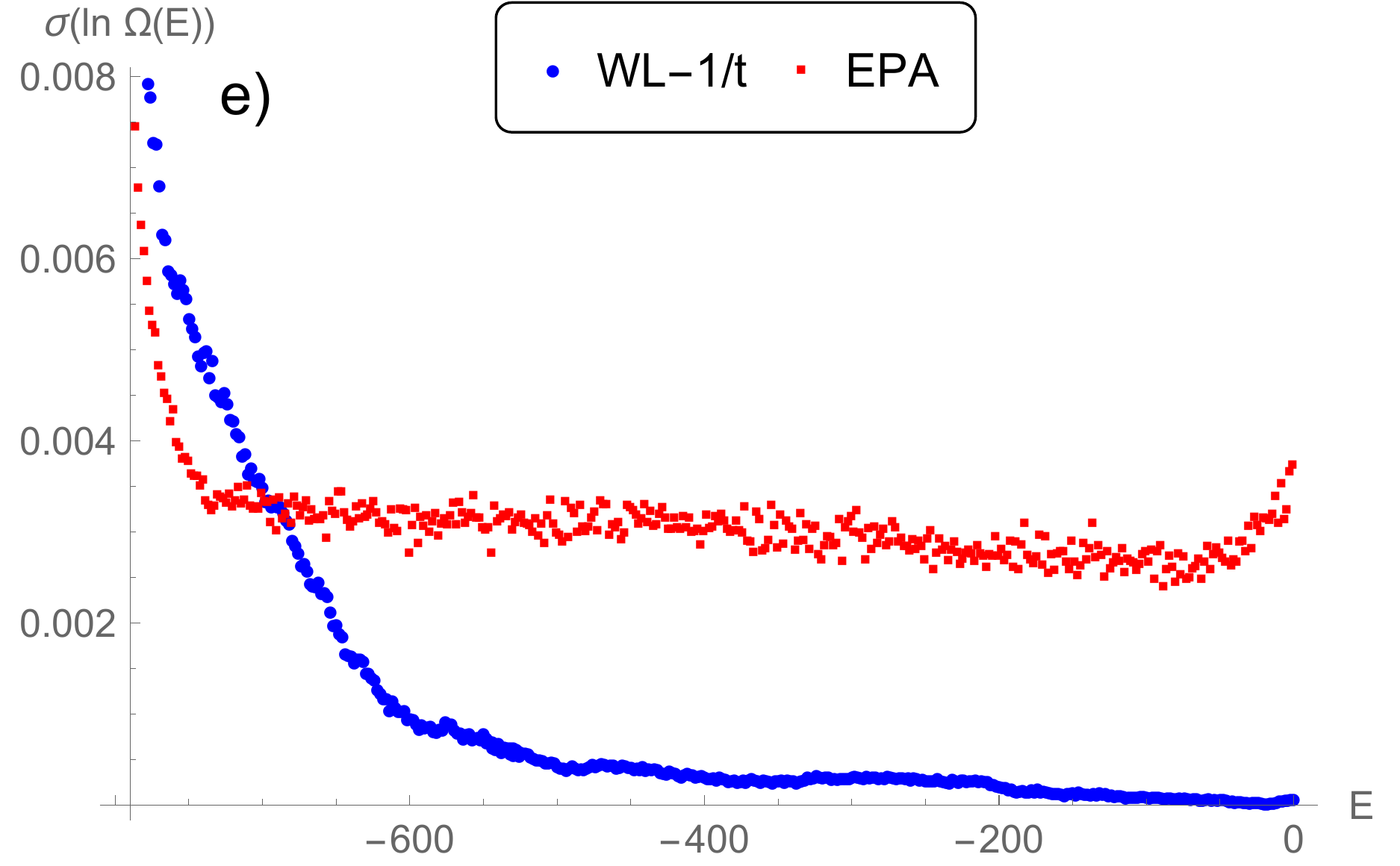}
\end{center}
\caption{Standard deviations of level entropies, $\sigma[\ln\Omega(E)]$, as sampled
  by the WL-$1/t$ and EPA approaches for a single sample from each hardness class,
  $k=3$ (a), $k=4$ (b), $k=5$ (c), $k=6$ (d), and $k=7$ (e).  WL-$1/t$ was performed
  with $4.8\times 10^{11}$ spin flip attempts for all samples, restricting the walk
  to energies $E\le E_\mathrm{max}$, where $E_\mathrm{max}=-500$ (the ground-state
  energy for these samples is roughly $E_0\approx -800$).  A pre-run of
  $2\times 10^{11}$ spin-flip attempts was performed to discover the range of
  possible energies for each sample; the main runs were started in the lowest-energy
  state found in the pre-run. All runs completed the first phase of the simulation
  here. For the EPA runs we used $R=10^6$, $\theta=10$, $\alpha^\ast = 0.86$,
  corresponding to $N_\beta \approx 100$ temperature steps and $4.8\times 10^{11}$
  spin flip attempts.}
  \label{fig:C}
\end{figure*}

While EPA provides the DOS with its absolute normalization, this is not natively the
case for WL. For a fair comparison hence different possible normalizations should be
considered. Two main normalization schemes are known. The first one consists of fixing the value of
$\Omega(E^\ast)$ at a specific energy, for example at the ground state $E^\ast = E_0$
or at $E^\ast = 0$. Alternatively, one can use the fact that the total number of
states is $2^N$, i.e.,
\begin{equation}
  \sum_E \Omega(E) = 2^N.
  \label{eq:total-number-of-states}
\end{equation}
In Fig.~\ref{fig:D} we show the effect of different normalizations applied to the
DOS estimate resulting from an EPA simulation (top panel) and a
WL-$1/t$ run (bottom panel) for a single realization (different realizations provide
comparable relative performances). It is apparent that different normalizations lead
to rather different statistical fluctuations for different energies, and that EPA and
WL-$1/t$ behave quite differently in this respect. Clearly, fixing the DOS at a
specific $E^\ast$ leads to zero fluctuations at this point. Averaged over all
energies, however, it is found that the normalization by the total number of states
leads to the lowest fluctuations, and for EPA these are very similar to those found
from the intrinsic normalization of EPA. It is also possible to consider entropy
{\em differences\/} such as $\ln \Omega(E)-\ln \Omega(E+2)$ that are intrinsically
normalization independent and also feature relatively low levels of fluctuation.  Two
of the curves are almost identical in the bottom panel in Fig.~\ref{fig:D}, because
the DOS has its largest value at $E=0$, hence the degeneracy is most accurately
estimated by WL-$1/t$ at $E=0$.

\section{Precision in different energy ranges}
\label{sec:energy-dependence}

A closer comparison of WL-$1/t$ and EPA is possible by considering the achieved
precision with the same computational effort, but resolved by energies. To this end,
we studied the behavior of $\sigma[\ln\Omega(E)]$, the standard deviation of the
level entropies. Here, we used the normalization to the total number of states
according to Eq.~\eqref{eq:total-number-of-states}. Simulations were performed with
both methods and the same set of samples, using parameters that ensure that the same
number of spin flips is performed. Figure~\ref{fig:C} shows the result of one sample
of each hardness class $k=3$, $4$, $5$, $6$, and $7$. While both methods have most
problems for the immediate vicinity of the ground state, as expected, one can see
that EPA results in a relatively small $\sigma$ for low energies, especially for hard
samples, and the results deteriorate only in the immediate vicinity of the ground
state. In contrast, the results from WL-$1/t$ deteriorate already for
$E \lesssim -600$, but it yields smaller fluctuations for higher energies, with the
best performance at $E\approx 0$. This seems plausible as WL spends much more time at
the higher energies with much larger entropies, whereas in PA the population is
continuously moved from high to low energies.

\section{Hardness of the planted ensemble}
\label{appendix:rhot_planted}

\begin{figure}[tb!]
\begin{center}
  \includegraphics[width=0.7\textwidth]{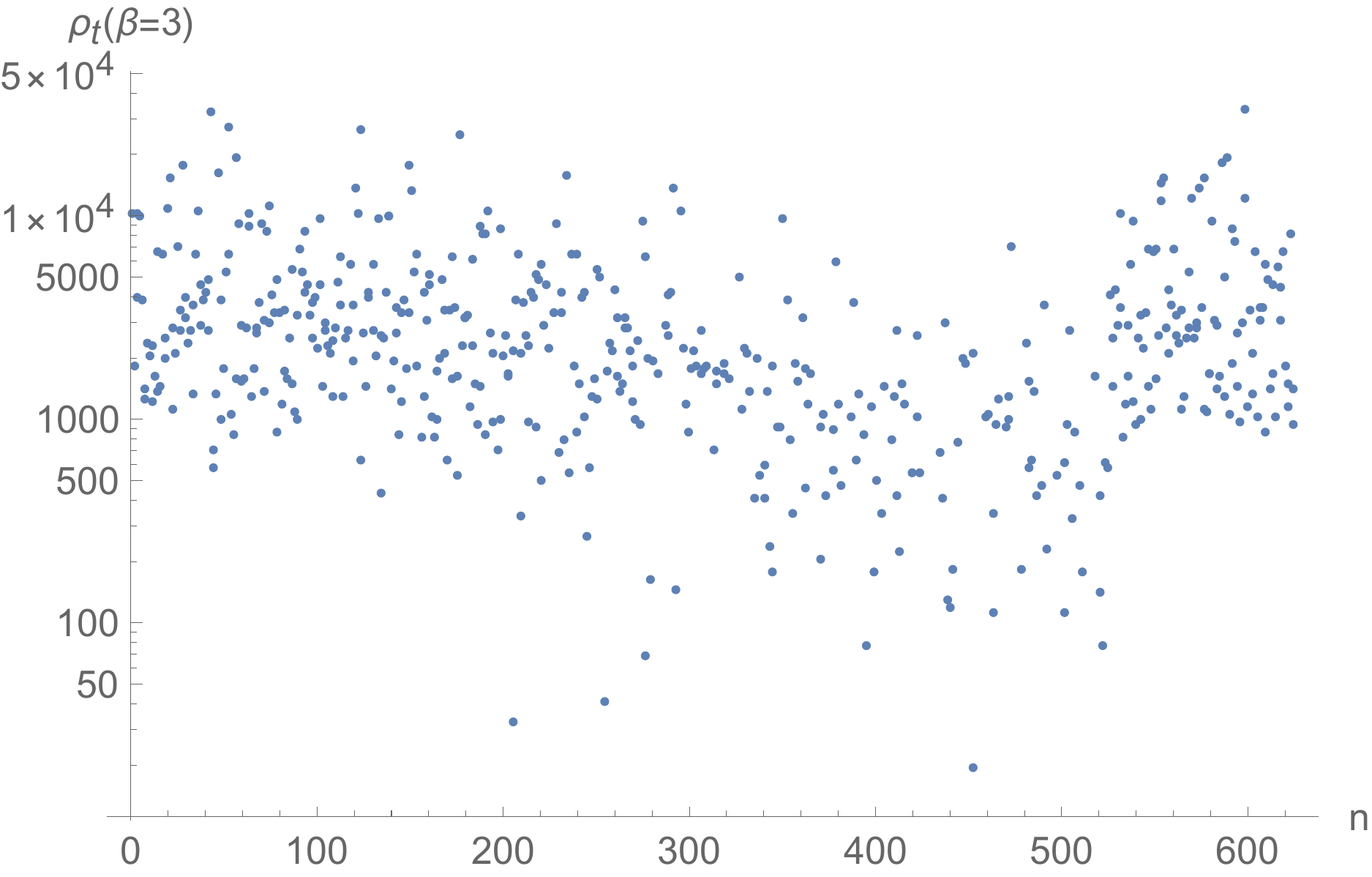}
\end{center}
  \caption{Mean-square family size $\rho_t$ at the lowest temperature
    in EPA for the 625 planted samples on Chimera graphs (see Sec.~\ref{sec:planted}). }
  \label{rhot_planted}
\end{figure}

It is interesting to consider some of the hardness measures for the EPA method also
for the other ensembles discussed here. Figure~\ref{rhot_planted} shows the
mean-square family size $\rho_t$ at the lowest temperature in EPA for the 625 planted
samples on Chimera graphs (see Sec.~\ref{sec:planted}).  The average value is
$\rho_t \approx 2000$, so the planted samples of this type are much harder then the
random ones (see Sec.~\ref{sec:varying_hardness}).

\end{document}